\documentclass[a4paper]{article}
\usepackage[T1]{fontenc}
\usepackage[utf8]{inputenc}
\usepackage{textcomp}
\usepackage{mathrsfs}
\usepackage{url}
\usepackage{bm}
\usepackage{amsmath}
\usepackage{amsthm}
\usepackage{amssymb}
\usepackage{stmaryrd}
\usepackage{graphicx}
\usepackage{esint}
\usepackage[authoryear]{natbib}
\PassOptionsToPackage{normalem}{ulem}
\usepackage{ulem}
\usepackage[unicode=true,
 bookmarks=false,
 breaklinks=false,pdfborder={0 0 1},backref=section,colorlinks=false]
 {hyperref}

\makeatletter

\pdfpageheight\paperheight
\pdfpagewidth\paperwidth

\newcommand{\aff}[1]{$^{#1}$} 
\newcommand{\corresp}[1]{} 
\newcommand{\shorttitle}[1]{} 
\newcommand{\shortauthor}[1]{} 
\newcommand{\affiliation}[1]{\thanks{#1}} 

\usepackage{enumerate}

\newtheorem{lemma}{Lemma}\newtheorem{corollary}{Corollary}

\usepackage{upgreek}
\usepackage[mathscr]{euscript}

\usepackage{bm}

\newcommand{\muSI}{\upmu_0}

\renewcommand{\vec}[1]{\bm{#1}} 
\newcommand{\vsf}[1]{\bm{\mathsf{#1}}} 

\renewcommand{\d}{\mathrm{d}} 
\renewcommand{\ij}{i} 
\newcommand{\Ident}{\vsf{I}} 
 
\newcommand{\Pperp}{\vsf{P}_{\perp}} 

\newcommand{\defeq}{\stackrel{\text{def}}{=}}

\providecommand{\grad}{\bm{\nabla}} 
\providecommand{\divv}{\bm{\nabla\cdot}} 
\providecommand{\cross}{\bm{\times}} 
\providecommand{\curl}{\bm{\nabla}\cross} 
\providecommand{\dotv}{\bm{\cdot}} 

\newcommand{\esub}[1]{\vec{e}_{#1}}

\newcommand{\half}{\mbox{\small\(\frac{1}{2}\)} }
\newcommand{\const}{{\mathrm{const}}}

\newcommand{\LdenRx}{\mathcal{L}_\Omega^{\rm Rx}} 
\newcommand{\LdenQRx}{\mathcal{L}_\Omega^{\rm A}} 
\newcommand{\LdenQPr}{\mathcal{L}_\Omega^{\textrm{A}_{-}}} 
\newcommand{\LdenEXB}{\mathcal{L}_\Omega^{\rm C}} 
 \newcommand{\SQRx}{\mathscr{S}_\Omega^{\rm A}}

    \newcommand{\Ees}{\vec{E}'}

\newcommand{\epsF}{}
\newcommand{\epsRx}{{\varepsilon}_{\mathrm{Rx}}}
\newcommand{\epsI}{{\varepsilon}_{\mathrm{I}}}

\newcommand{\muOm}{\mu_\Omega }
\newcommand{\nuOm}{\nu_\Omega}
\newcommand{\uRx}{\vec{u}^{\rm Rx}}

\newcommand{\cIOL}{\vec{C}}
\newcommand{\lamIOL}{\vec{\lambda}}
\newcommand{\lamstar}{\bm{\lambda}_{*}}

\newcommand{\muIOL}{\mu^{\rm P}_{\Omega}}

\makeatother

\begin{document}
\shorttitle{RxMHD with IOL constraint}

\shortauthor{R. L. Dewar, Z. S. Qu \today} 

\title{Relaxed Magnetohydrodynamics with Ideal Ohm's Law Constraint (arxiv
v2)}
\author{R.~L. Dewar\aff{1} \corresp{\email{robert.dewar@anu.edu.au}}
\and Z.~S. Qu\aff{1}}

\maketitle
\affiliation{\aff{1}Mathematical Sciences Institute, The Australian
National University, Canberra, ACT 2601, Australia}




\begin{abstract}
The gap between a recently developed dynamical version of relaxed
magnetohydrodynamics (RxMHD) and ideal MHD (IMHD) is bridged by approximating
the zero-resistivity ``Ideal'' Ohm's Law (IOL) constraint using
an augmented Lagrangian method borrowed from optimization theory.
The augmentation combines a pointwise vector Lagrange multiplier method
and global penalty function method and can be used either for iterative
enforcement of the IOL to arbitrary accuracy, or for constructing
a continuous sequence of magnetofluid dynamics models running between
RxMHD (no IOL) and weak IMHD (IOL almost everywhere). This is illustrated
by deriving dispersion relations for linear waves on an MHD equilibrium. 
\end{abstract}

\section{Introduction}

\label{sec:Intro1}

\subsection{Basics\label{subsec:Basics}}

In this paper choosing \emph{constraint equations} is central to our
approach to developing new fluid models. The concept of a constraint
equation occurs in both the variational approach to classical mechanics
{[}see e.g. \citet{Goldstein_80}{]} and optimization theory {[}see
e.g. \citet{Nocedal_Wright_2006}{]}. While both traditionally treat
finite-dimensional systems, the language and techniques of these fields
can also help in understanding the infinite-dimensional dynamics of
non-dissipative continuous media. In the following we shall distinguish
between a \emph{hard constraint}, i.e. one that is enforced exactly,
a \emph{soft constraint}, one that is enforced only approximately,
and a \emph{weak} version of a hard constraint, one that is enforced
as the limiting case of a sequence of soft constraints (formulating
such a method being the goal of this work, which it is hoped will
lead to a physical regularization\footnote{We use \emph{regularization} in the physics sense --- adjusting for
incipient singular behaviour in a way that is consistent with physics
on scales outside the strict domain of applicability of a mathematical
model. This goes somewhat beyond the mathematical sense of adjusting
a problem to avoid ill-posedness.} of MHD that allows reconnection).

We also distinguish between \emph{microscopic}, i.e. acting within
each \emph{fluid element} or infinitesimal parcel of fluid, and \emph{macroscopic}
constraints, i.e. \emph{global} within a spatial domain $\Omega$
of the fluid (or subdomain if the system is partitioned into multiple
regions).

The mathematical model we seek to regularize is \emph{Ideal }MHD (IMHD),
a special case in the general field of magnetohydrodynamics (MHD).
In the general, resistive case Ohm's Law is $\Ees=\eta\vec{j}$, where
\begin{equation}
\Ees[\vec{u}]\defeq\vec{E}+\vec{u}\cross\vec{B}\label{eq:advectedE}
\end{equation}
is the electric field observed in the local frame of each fluid element,
$\vec{E}$ being the electric field in the lab frame. These elements
are advected in the fluid velocity field $\vec{u}(\vec{x},t)$ (i.e.
$\dot{\vec{x}}=\vec{u}$ at each spatial point $\vec{x}$ and time
$t$). Also $\vec{B}(\vec{x},t)$ is the magnetic field, $\eta$ is
the resistivity and $\vec{j(\vec{x},t)}$ is the electric current
density (N.B. $\vec{j}=\curl\vec{B}/\muSI$ in standard non-relativistic
MHD, where $\muSI$ is the vacuum permeability constant used in SI
electromagnetic units). We have exhibited $\vec{u}$ as an explicit
argument for use later in the paper, while leaving dependencies on
$\vec{x},t,\vec{E,}$ and $\vec{B}$ implicit.

To get IMHD, set $\eta=0$ so that $\Ees=0$, giving what is often
called the \emph{Ideal Ohm's Law} (IOL): 
\begin{equation}
\vec{E}+\vec{u}\cross\vec{B}=0\;.\label{eq:idealOhm}
\end{equation}

While $\vec{E}$ is not usually explicit in the IMHD equations, this
is only because it is eliminated between \eqref{eq:idealOhm}, after
taking the curl of both sides, and the \emph{Maxwell--Faraday induction
equation} 
\begin{equation}
\curl\vec{E}=-\epsF\partial_{t}\vec{B}\;,\label{eq:MaxFad}
\end{equation}
to give the IMHD magnetic-field propagation equation 
\begin{equation}
\epsF\partial_{t}\vec{B}+\curl(\vec{u}\cross\vec{B})=0\;.\label{eq:dBdt}
\end{equation}

With the ``pre Maxwell'' Ampère's Law $\vec{j}=\curl\vec{B}/\muSI$
and $\divv\vec{B}=0$, the Maxwell-Faraday equation \eqref{eq:MaxFad}
plays the important role of preserving Galilean invariance {[}\citet[Sec.~5.4]{Hosking_Dewar_2015};
\citet{Webb_Anco_2019}{]}, independent of whether or not the IOL
equation is enforced. Thus we shall retain it in the following development
of a dynamical relaxation theory.

Equation \eqref{eq:MaxFad} can be viewed as a holonomic constraint
on $\vec{E}$, and likewise $\divv\vec{B}=0$ is a holonomic constraint
on $\vec{B}$, i.e. we can remove these constraints from consideration
by expressing the constrained variables in terms of fewer unconstrained
variables. Here these are the vector and scalar potentials $\vec{A}$
and $\Phi$, respectively, in terms of which 
\begin{align}
\vec{B} & =\curl\vec{A}\;,\label{eq:Brep}\\
\vec{E} & =-\epsF\partial_{t}\vec{A}-\grad\Phi\;.\label{eq:Erep}
\end{align}
These imply $\divv\vec{B}=0$ and also \eqref{eq:MaxFad}, as is easily
seen by calculating $\curl\vec{E}$.

We restrict the choice of gauge to be such that $\Phi$ is a spatially
\emph{single-valued} potential and such that $\partial_{t}\vec{A}=0$
in equilibrium cases in a frame (the LAB frame) where $\partial_{t}\,\cdot=0$,
so $\vec{A}$ has no effect on $\vec{E}$ in that static case. Of
course the vector potential still does play an explicit role in describing
plasma equilibria because the magnetic flux threading a loop is $\oint\vec{A}\dotv\d\vec{l}$.
Dynamically, only $\partial_{t}\vec{A}$ contributes to inductive
e.m.f.s $\oint\vec{E}\dotv\d\vec{l}$ around closed loops. In our
case, we assume e.m.f.s are \emph{zero} around any loop on the \emph{boundary}
$\partial\Omega$ --- the trapped-flux boundary condition of RxMHD
{[}see Appendix B of \citet{Dewar_Yoshida_Bhattacharjee_Hudson_2015}{]}.
Aside from this restriction, there is still considerable gauge freedom
in $\vec{A}$. If we choose Coulomb gauge, $\divv\vec{A}=0$, the
potential representation is an example of the Helmholtz decomposition
of an arbitrary vector field into the sum of curl-free and divergence-free
vector fields, but we shall not make this gauge choice except in Sections~\ref{subsec:PhiVariation}
and \ref{sec:WKBRxMHD} --- we shall treat the magnetic helicity
term carefully in our general derivation of the conservation form
momentum equation in order to make it gauge invariant.

\subsection{Methodology: Variational principles and Euler--Lagrange equations\label{subsec:VarPrinciples-ELeqns}}

In mechanics and optimization theory there are objective functions
whose extrema --- maxima, minima and saddle points --- are given
by \emph{Euler--Lagrange} (EL) equations, which are found by setting
first derivatives of these functions to zero. In mechanics such functions
are \emph{Hamiltonians} whose extrema give stable or unstable equilibria,
or \emph{actions, }time integrals of \emph{Lagrangians}, whose extrema
give physical time evolution equations (Hamilton's Principle).

The main aim of this paper is to use an infinite-dimensional generalization
of Hamilton's Principle in which partial derivatives are replaced
by\emph{ functional }derivatives {[}see e.g. \citet{Morrison_98}{]}
of action integrals incorporating the IOL constraint, and also global
entropy, magnetic-helicity and cross-helicity constraints. These functional
derivatives are with respect to the basic physical fields, e.g. $\Phi$,
$\vec{A}$, $\vec{u}$, etc., describing the state of the system and
are set to zero to find a set of Euler--Lagrange equations which
together are sufficient to describe the dynamics of the system. For
brevity we shall refer e.g. to the equation found by setting the functional
derivative with respect to $\Phi$ as the ``$\delta\Phi$-EL equation''.

\subsection{Relaxation\label{subsec:Relaxation}}

See Appendix~\ref{sec:RxHistory} for a brief history of the variational
approach to finding relaxed plasma equilibrium states by minimizing
the IMHD energy functional using one or more IMHD invariants as global
constraints. This construction implies immediately that such relaxed
magnetostatic states are a special \emph{subset} of all possible IMHD
equilibria, most of which, being of higher energy, are likely to be
more unstable than relaxed states.

In this paper we instead seek to find a time-dependent variational
formulation for relaxed plasma systems going through a \emph{dynamical}
phase as they transition from one equilibrium state to another (e.g.
due to boundary deformations). Thus, instead of minimizing energy,
we use Hamilton's variational Principle, widely regarded as the most
fundamental principal in all mathematical physics, from general relativity
through classical mechanics to quantum field theories (for instance
connecting symmetries and conservation laws by Noether's theorem).
As we are attempting to establish a new classical field theory related
to, but different from, ideal magnetohydrodynamics (IMHD), it is appropriate
to seek new magnetofluid models by modifying the IMHD Hamilton's Principle.

Following this precept, \citet{Dewar_Burby_Qu_Sato_Hole_2020} derived
a new dynamical magnetofluid model, \uline{R}ela\uline{x}ed
\uline{M}agneto\uline{H}ydro\uline{D}ynamics (RxMHD), from
Hamilton's Action Principle using a phase-space version of the magnetofluid
Lagrangian with a noncanonical momentum field $\vec{u},$ physically
identified as the lab-frame mass-flow velocity, and a kinematically
constrained velocity field $\vec{v}$ (the fluid velocity relative
to a magnetic-field-aligned flow). The resulting Euler--Lagrange
equations generalize from statics to dynamics the usual relaxation-by-energy-minimization
concept developed by \citet{Taylor_86} for flowless plasma equilibria,
and its generalization to equilibria with steady flow by various authors:
\citet{Finn_Antonsen_83,Hameiri_98,Vladimirov_Moffatt_Ilin_99,Hameiri_2014,Dennis_Hudson_Dewar_Hole_2014a}.
These generalized Taylor equilibria were shown by \citet{Dewar_Burby_Qu_Sato_Hole_2020}
to be consistent with RxMHD when time derivatives are set to zero.
However, specific cases of equilibria with flows not aligned with
the magnetic field have been limited to axisymmetric equilibria, whereas
in this paper we aim to treat more general, non-axisymmetric (3-D)
equilibria with flow, as well as time-dependent problems such as the
calculation of the spectrum of normal modes of oscillation of 3-D
relaxed equilibria.

The advection equation for $\vec{B}$, \eqref{eq:dBdt}, implies the
``frozen-in flux constraint'', which, as discussed by \citet{Newcomb_58},
preserves the topology of magnetic field lines. This prevents field-line
breaking and \emph{reconnection} from forming new structures, such
as magnetic islands, and this frustration of topological changes leads
to singularities developing as time tends toward infinity \citet{Grad_67}.

Though in this paper we proceed in a formal way by simply inserting
constancy constraints of selected IMHD invariants as postulates, historically
the heuristic assumption motivating relaxation theory is that, if
it would be energetically favourable to do so, and on a long enough
timescale, ``nature will find a way'' for reconnection to occur,
either due to the magnifying effect of large gradients on small but
finite resistivity at singularities, or through ``anomalous'' phenomena
such as turbulence. Thus in the RxMHD of \citet{Dewar_Burby_Qu_Sato_Hole_2020}
the continuum of local frozen-in flux constraints is replaced by only
two constraints involving $\vec{B}$, the two global IMHD invariants
magnetic helicity and cross helicity.

However, as will be argued in Subsection~\ref{subsec:EqmPhi}, there
is reason to believe that, for general three-dimensional equilibria
with \emph{non-integrable} magnetic field dynamics, imposing \eqref{eq:idealOhm}
as a hard constraint would lead to an ill-posed variational principle
with no smooth extremum. In this case we regularize the problem by
approaching an IOL-constrained state through a sequence of softly
constrained states where the IOL constraint is not exactly satisfied.

For a dynamical relaxed MHD theory to be fully satisfactory we require
it to be well-posed mathematically and desire it to agree with \emph{ideal}
MHD in two cases: (i) on the boundary $\partial\Omega(t)$, because
MRxMHD interfaces are regarded as arbitrarily thin sheets of IMHD
fluid; and (ii) in an equilibrium state with steady flow, when one
imagines any transient non-ideal behaviour to have died away, justifying
the \emph{Principle of IMHD-Equilibrium Consistency} {[}\citet{Dewar_Burby_Qu_Sato_Hole_2020}{]}.

This Consistency Principle was satisfied by the one flowing equilibrium
test case \citet{Dewar_Burby_Qu_Sato_Hole_2020} looked at using their
RxMHD formulation, the rigidly rotating axisymmetric steady-flow equilibrium.
However RxMHD does not \emph{enforce} the IOL constraint \eqref{eq:idealOhm},
so there is no reason to believe that ideal consistency would necessarily
apply to more general relaxed equilibria. {[}Indeed, \citet{Dewar_Burby_Qu_Sato_Hole_2020}
showed that small dynamical perturbations about an equilibrium exhibited
no tendency to preserve the IOL constraint.{]}

Specifically, we are interested in\emph{ non}-axisymmetric relaxed
steady-flow toroidal equilibria such as may occur in stellarators.
The elliptic nature of RxMHD (when flows are small) makes it reasonable
to assume that smooth solutions of the RxMHD equations exist for such
equilibria. We argue in Subsection~\ref{subsec:EqmPhi} that, generically,
magnetic field and fluid flow lines on these smooth RxMHD solutions
will be chaotic so their ergodic properties will exhibit complexity
on all scales.

While RxMHD offers no impediment to the formation of such fractal
structure {[}one of the principal motivations for the develoment of
the SPEC code, \citet{Hudson_etal_2012b}{]} the same is \emph{not}
true for IMHD where the topological constraints arising from its frozen-in-flux
properties (see above) force the formation of singularities. The ability
of the SPEC equilibrium code to study difficult physical problems
{[}\citet{Qu_etal_2020}{]}, and subtle fundamental problems involving
chaos {[}\citet{Qu_Hudson_Dewar_Loizu_Hole_2021}{]}, motivates our
current endeavour to extend the RxMHD formalism on which it is based
to make it closer to IMHD but to retain sufficient topological relaxation
to allow magnetic island formation and chaos, thus allowing further
extension of SPEC to hande time-dependent problems in three-dimensional
geometries.

\subsection{Background flow\label{subsec:Background-flow}}

We define a \emph{fully relaxed} RxMHD equilibrium as one where the
electrostatic potential $\Phi$ has relaxed to a constant value throughout
a volume $\Omega$, so $\vec{E}=0$. As \citet{Finn_Antonsen_83}
recognized, this would occur in the extreme case where magnetic field
lines fill $\Omega$ ergodically, because dotting both sides of \eqref{eq:idealOhm}
with $\vec{B}$ gives the derivative along $\vec{B}$ as $\vec{B}\dotv\grad\Phi=0$.
As we shall see, constant $\Phi$ implies purely parallel flow, $\vec{u}=\vec{u}_{\parallel}$,
whose magnitude is constrained by the steady-flow continuity equation
$\divv[(\rho/B)\vec{B}u_{\parallel}]=\vec{B}\dotv\grad(\rho u_{\parallel}/B)=0$,
where $\rho$ is mass density. For consistency again with the (unachievable)
fully ergodic limit, we define fully relaxed parallel flow as such
that $\rho u_{\parallel}/B=\const$. We denote this special parallel
flow velocity as 
\begin{equation}
\uRx\defeq\frac{\nuOm\vec{B}}{\muSI\rho}\;,\label{eq:ufullRx}
\end{equation}
where $\nuOm$ is a constant throughout $\Omega$ --- its significance
in the RxMHD formalism is explained below:

In the variational $\vec{u},\!\vec{v}$ dynamical relaxation formalism
of \citet{Dewar_Burby_Qu_Sato_Hole_2020}, EL equations for $\vec{u}$,
$\vec{v}$, $\vec{B}$ and pressure $p$ are derived variationally
from Hamilton's Principle, while the mass continuity equation is built
in as a holonomic constraint. The fully relaxed flow $\uRx$ occurs
in these EL equations, with $\nuOm$ arising as the Lagrange multiplier
for the magnetic-helicity constraint in the phase-space Lagrangian.
Specifically, the EL equation arising from free variations of $\vec{u}$
is 
\begin{equation}
\vec{u}=\uRx+\vec{v}\;,\label{eq:vXshift}
\end{equation}
so $\vec{v}$ is the\emph{ relative flow}, the fluid velocity relative
to the fully relaxed flow velocity $\uRx$.

Noting from \eqref{eq:ufullRx} that $\divv(\rho\uRx)=0$, we see
that $\divv(\rho\vec{u})=\divv(\rho\vec{v})$. Thus the continuity
equation holds for both $\vec{u}$ and $\vec{v}$, i.e. both flows
are microscopically mass-conserving. Also, $\vec{u}\cross\vec{B}=\vec{v}\cross\vec{B}$,
so $\Ees[\vec{u}]=\Ees[\vec{v}]$. In order to preserve \eqref{eq:vXshift}
in the variational formulation (see later), the version of the IOL
constraint we shall be using in the body of this paper is $\Ees[\vec{v}]=0$,
which becomes equivalent to $\Ees[\vec{u}]=0$ only \emph{after} the
Euler--Lagrange equations are derived.

\subsection{Domains and boundaries\label{subsec:Domains-and-boundaries}}

For most purposes in this paper it is sufficient to restrict attention
to plasma within a \emph{single} domain $\Omega(t)$ that is closed,
of genus at least 1, and whose boundary $\partial\Omega(t)$ is smooth,
gapless, perfectly conducting and time-dependent. However we note
this is part of a larger project, the development of Multiregion Relaxed
MHD (MRxMHD) \citet{Dewar_Yoshida_Bhattacharjee_Hudson_2015}, in
which $\Omega$ is but a subregion of a larger plasma region, partitioned
into multiple relaxation domains physically separated by moving \emph{interfaces}.
As $\partial\Omega(t)$ is the union of the inward-facing sides of
the interfaces $\Omega(t)$ shares with its neighbours, it transmits
external forcing to the restricted subsystem within $\Omega(t)$ and
imparts equal and opposite reaction forces on the neighbouring subdomains.

We take the interfaces to be perfectly flexible and impervious to
mass and heat transport. We also take them to be impervious to magnetic
flux like a superconductor, implying the \emph{tangentiality condition}
\begin{equation}
\vec{n}\dotv\vec{B}=0\:\:\text{on}\:\partial\Omega\;,\label{eq:tangential}
\end{equation}
where $\vec{B}\defeq\curl\vec{A}$ is the magnetic field and $\vec{n}$
is a unit normal at each point on $\partial\Omega$ (here and henceforth
leaving the argument $t$ implicit in $\Omega$, $\vec{n}$ etc.).
Also, to conserve magnetic fluxes trapped within $\Omega$, loop integrals
of the vector potential $\vec{A}$ within the interfaces must be conserved
{[}see e.g. \citet{Dewar_Yoshida_Bhattacharjee_Hudson_2015}{]}.

\subsection{Layout of this paper }

The phase-space Lagrangian variational approach to deriving ideal
MHD equations is briefly reviewed in Section~\ref{sec:IMHDPSL},
then some general implications of the IOL when it is a hard constraint\textbackslash{}
are discussed in Section~\ref{sec:ModOhm} including speculations
in Subsection~ \ref{subsec:EqmPhi} on the implications of chaos
and ergodic theory on flows in three-dimensional systems, in Subsection~\ref{subsec:EXBdrift}
the $\vec{\ensuremath{E}}\cross\vec{B}$ drift is derived.

In Section~\ref{sec:Constraints} the adaptation of the augmented
Lagrangian penalty function method from optimization theory to the
physical purpose of approximating the IOL constraint is discussed
as a softly constrained optimization problem in Subsction~\ref{subsec:SoftIOL},
and the specific Lagrangian density constraint term for this method
is given in Subsection~\ref{subsec:AugLag}. The entropy, magnetic
helicity and cross-helicity conservation constraints used in Relaxed
MHD theory are discussed in Subsection~\ref{subsec:Invariants},
and the complete phase space Lagrangian to be used in this paper is
constructed in Subsection~\ref{subsec:macroPSL}.

In Section~\ref{sec:RxMHDEL} the Euler--Lagrange equations, including
an equation of motion in momentum conservation form, are derived formally
in Subsection~\ref{subsec:QRxMHDSvar}, and in specific forms in
Subsections~\ref{subsec:Conservation-form}--\ref{subsec:Lagrangian-variation}
where the IOL constraint term provides new contributions that vanish
only when the constraint is satisfied. In addition to the momentum
equation form, an equation of motion in Bernoulli form is derived.
A physical interpretation of the Lagrange multiplier for the IOL constraint
in terms of a polarization field is also mentioned.

Section~\ref{sec:WKBRxMHD} illustrates the implementaton of the
augmented Lagrangian method for linear waves propagating on an IOL-compliant
equilibrium in the WKB approximation. A continuous family of dispersion
relations for wave residuals $\widetilde{\cIOL}$ ranging from zero
in the IMHD case to its value in the RxMHD case, where it is the perturbed
Lagrange multiplier $\widetilde{\lamIOL}$ that is set to zero.

The Conclusion, Section~\ref{sec:Concl}, briefly summarizes what
has been achieved in this paper and what more needs to be done. More
detail on derivations of equations is available as online Supplementary
Material in an unabridged version of this paper.

A brief historical overview of MHD relaxation theory is given in Appendix~\ref{sec:RxHistory}
and some useful vector and dyadic calculus identities are derived
in Appendix~\ref{sec:lemmas}, in particular the little-known identity
\eqref{eq:LemgradBdotf}, which is crucial for getting the general
form of the momentum equation \eqref{eq:xiEL} into a general conservation
form, \eqref{eq:MomConsvnForm}.

\section{Ideal MHD in phase space\label{sec:IMHDPSL}}

The mathematical foundation on which our dynamical relaxation formalism
is built is a noncanical form (which we call the $\vec{u}$, $\vec{\ensuremath{v}}$
picture) of the canonical MHD Hamiltonian, and a \emph{Phase-Space
Lagrangian} (PSL). Here we review how Hamilton's action principle
leads to IMHD when microscopic constraints on entropy and magnetic
flux are applied. Later we show how RxMHD arises when these are replaced
by global constraints using the same PSL formalism.

Both ideal and relaxed MHD starts from the canonical MHD Hamiltonian

\begin{align}
H^{{\rm MHD}}[\vec{x},\bm{\pi},t] & =\int_{\Omega}\mathcal{H^{{\rm MHD}}}\,\d V\;,\nonumber \\
\text{with}\:\:\mathcal{H^{{\rm MHD}}}(\vec{x},\bm{\pi},t) & \defeq\frac{\bm{\pi}^{2}}{2\rho}+\frac{p}{\gamma-1}+\frac{B^{2}}{2\muSI}\;,\label{eq:Hgen}
\end{align}
where $\bm{\pi}\left(\vec{x},t\right)$ is the canonical momentum
density, the analogue of $p$ in finite-dimensional classical dynamics.

The analogue of $q$ is not $\vec{x}$ the Eulerian independent variable
but $\vec{r}$, the Lagrangian position with respect to a given reference
frame. We do not make this explicit as we shall always work in the
Eulerian picture, but the Lagrangian picture in the background \emph{does}
manifest in interpreting variations. {[}This is discussed in more
detail by \citet{Dewar_Burby_Qu_Sato_Hole_2020}.{]} For instance,
the analogue of the variation $\delta q$ at fixed $t$ is $\Delta\vec{x}=\bm{\xi}(\vec{x},t)$,
\emph{the Lagrangian fluid displacement in Eulerian representation,}
and the analogue of the variation $\dot{q}\,\delta t$ is $\vec{v}(\vec{x},t)\,\delta t$,
which we shall refer to as the Lagrangian velocity field (not always
the same as the Eulerian velocity $\vec{u}$). Both ideal and RxMHD
also use the constrained \emph{kinematic variation} \citet{Newcomb_62},
\begin{equation}
\delta\vec{v}=\partial_{t}\bm{\xi}+\vec{v}\dotv\grad\bm{\xi}-\bm{\xi}\dotv\grad\vec{v}\;.\label{eq:vvar}
\end{equation}

They also use the \emph{mass density variation} 
\begin{equation}
\delta\rho=-\divv(\rho\,\bm{\xi})=-\rho\divv\bm{\xi}-\bm{\xi}\dotv\grad\rho\;,\label{eq:rhovar}
\end{equation}
which is an expression of the \emph{microscopic} conservation of mass
and can be found by integrating the perturbed continuity equation
\begin{equation}
\partial_{t}\rho+\divv(\rho\vec{v})=0\:\:\Leftrightarrow\:\:\frac{\d\rho}{\d t}=-\rho\divv\vec{v}\label{eq:Continuity}
\end{equation}
along varied Lagrangian trajectories $\vec{r}(t|\vec{x}_{0})$ {[}\citet{Frieman_Rotenberg_60}{]}
and expressing this Lagrangian variation in the Eulerian picture {[}\citet{Newcomb_62}{]}.

Instead of seeking a Poisson bracket to get phase-space dynamics from
$H$ {[}see e.g. \citet{Morrison_98}{]}, we instead work directly
with the canonical \emph{phase-space Lagrangian} (PSL) density $\mathcal{L^{{\rm MHD}}}$,
\begin{align}
L_{{\rm ph}}^{{\rm MHD}}[\vec{x},\vec{v},\bm{\pi}] & =\int_{\Omega}\!\!\left[\bm{\pi}\dotv\vec{v}-\mathcal{H^{{\rm MHD}}}(\vec{x},\bm{\pi},t)\right]\,\d V\;,\nonumber \\
 & \defeq\int_{\Omega}\!\!\,\mathcal{L^{{\rm MHD}}}(\vec{x},\bm{\pi},t)\,\d V\nonumber \\
 & =\int_{\Omega}\!\!\left(\bm{\pi}\dotv\vec{v}-\frac{\bm{\pi}^{2}}{2\rho}-\frac{p}{\gamma-1}-\frac{B^{2}}{2\muSI}\right)\,\d V\;,\label{eq:microPSL}
\end{align}
and the corresponding canonical \emph{phase-space action}, 
\begin{equation}
\mathscr{S_{{\rm ph}}^{{\rm MHD}}}\defeq\iint_{\Omega}\!\!\mathcal{L^{{\rm MHD}}}\,\d V\d t\label{eq:PSA}
\end{equation}
as the primary tools, deriving Euler--Lagrange (EL) equations from
\emph{Hamilton's Principle of stationary action}, 
\begin{equation}
\delta\mathscr{S_{{\rm ph}}^{{\rm MHD}}}=0\;,\label{eq:PSHamPrinc}
\end{equation}
varying phase space paths under appropriate constraints.

We have used the subscript notation $\cdot_{{\rm ph}}$ on the Lagrangian
$L_{{\rm ph}}$ and the action $\mathscr{S}_{{\rm ph}}$ to make it
clear the PSL defined in \eqref{eq:microPSL} is fundamentally different
from the more usual configuration space Lagrangian and action. This
is because, in \eqref{eq:PSA}, $\bm{\pi}$ is now regarded as \emph{freely
variable}, so the dimensionality of the space of allowed variations
is doubled in the phase-space action principle.

For instance, varying $\bm{\pi}$ in \eqref{eq:microPSL} gives the
$\delta\bm{\pi}$-EL equation $\delta\mathscr{S}_{{\rm ph}}/\delta\bm{\pi}=\vec{v}-\bm{\pi}/\rho=0$,
i.e., multiplying by $\rho$, the analogue of $p=m\dot{q}$ is seen
to be $\bm{\pi}=\rho\vec{v}$, as expected. Likewise, using the microscopic
holonomic constraints of entropy and flux, $\delta p=-\gamma p\divv\bm{\xi}-\bm{\xi}\dotv\grad p$
and $\delta\vec{B}=\curl(\bm{\xi}\cross\vec{B})$, respectively, one
can verify that the Euler--Lagrange equation arising from Lagrange-varying
$\vec{x}$ (i.e. varying $\bm{\xi}$) is just the IMHD equation of
motion.

However, as it is not customary in fluid mechanics to work with canonical
momenta, we follow \citet{Burby_2017} in exploiting the freedom afforded
by the PSL to work with a velocity-like phase-space variable $\vec{u}$,
obtained by the \emph{noncanonical} change of variable $\bm{\pi}=\rho\vec{u}$.
Then the canonical Hamiltonian density \eqref{eq:Hgen} becomes the
noncanonical Hamiltonian density 
\begin{align}
\mathcal{H}_{{\rm nc}}^{{\rm MHD}}(\vec{x},\vec{u},t) & =\frac{\rho\vec{u}^{2}}{2}+\frac{p}{\gamma-1}+\frac{B^{2}}{2\muSI}\;,
\label{eq:HMHD}
\end{align}
and the PSL density in noncanonical form becomes, from \eqref{eq:microPSL},
\begin{align}
\mathcal{L}_{{\rm nc}}^{{\rm MHD}}(\vec{x},\vec{u},\vec{v},t) & \defeq\rho\vec{u}\dotv\vec{v}-\mathcal{H}_{{\rm nc}}^{{\rm MHD}}(\vec{x},\vec{u},t)\nonumber \\
 & =\rho\vec{u}\dotv\vec{v}-\frac{\rho\vec{u}^{2}}{2}-\frac{p}{\gamma-1}-\frac{B^{2}}{2\muSI}\;.\label{eq:PSLnc}
\end{align}
As neither $p$ nor $\vec{B}$ depends on $\vec{u}$, the $\delta\vec{u}$-EL
equation is $\delta\mathscr{S}_{{\rm ph}}^{{\rm MHD}}/\delta\vec{u}=\rho\vec{v}-\rho\vec{u}=0$,
i.e. in IMHD we have $\vec{v}=\vec{u}$. The IMHD equation of motion,
which can be written in conservation form as

\begin{equation}
\partial_{t}(\rho\vec{u})+\divv\vsf T_{{\rm {\rm MHD}}}=0\;,\label{eq:IMHDmomeqn}
\end{equation}
where 
\begin{equation}
\vsf T_{{\rm {\rm MHD}}}\defeq\rho\vec{u}\vec{u}+\left(p+\frac{B^{2}}{2\muSI}\right)\Ident-\frac{\vec{B}\vec{B}}{\muSI}\;,\label{eq:IMHD}
\end{equation}
follows as it does for the $\bm{\xi}$-EL equation in the canonical
form.

It will shown below that an isothermal version of IMHD can be derived
by replacing the holonomic variational constraint $\delta p=-\gamma p\divv\bm{\xi}-\bm{\xi}\dotv\grad p$
with a \emph{global} entropy conservation constraint, giving thermal
relaxation, a more realistic model for hot plasmas than the microscopic
entropy constraints implied by $\delta p=-\gamma p\divv\bm{\xi}-\bm{\xi}\dotv\grad p$.

\section{Implications of the IOL constraint \label{sec:ModOhm}}

In this section we examine consequences of applying the IOL constraint
\eqref{eq:idealOhm} in the form (see Subsection~\ref{subsec:Background-flow})
$\Ees[\vec{v}]=0$, which can be written $-\vec{E}=\vec{v}\cross\vec{B}$
or $\epsF\partial_{t}\vec{A}+\grad\Phi=\vec{v}\cross\vec{B}$.

\subsection{$\vec{E}\cross\vec{B}$ drift \label{subsec:EXBdrift}}

As $\Ees=\vec{E}+\vec{v}\cross\vec{B}$ we have the two identities
\begin{align}
\vec{B}\cross\Ees & =-\vec{E}\cross\vec{B}+(B^{2}\Ident-\vec{B}\vec{B})\dotv\vec{v}\;,\label{eq:Bcrosscond}\\
\text{and}\quad\vec{v}\cross\Ees & =-\vec{E}\cross\vec{v}-(v^{2}\Ident-\vec{v}\vec{v})\dotv\vec{B}\;.\label{eq:ucrosscond}
\end{align}

Equation \eqref{eq:Bcrosscond} leads to a decomposition of the relative
fluid flow into a component $v_{\parallel}$ tangential to $\vec{B}$
at $\vec{x}$, and a component $\vec{v}_{\perp}$, its projection
onto the plane transverse to $\vec{B}$, the ``$\vec{E}\cross\vec{B}$
drift,'' 
\begin{equation}
\vec{v}_{\perp}=\frac{\vec{E}_{\perp}\cross\vec{B}}{B^{2}}\;.\label{eq:QEXB}
\end{equation}

It is usually safe to assume $B^{2}\neq0$ anywhere in toroidally
confined plasmas, so the representation \eqref{eq:QEXB} generally
applies everywhere, and to both equilibrium and dynamic ES MHD cases.\footnote{The case $u^{2}=0$ may well occur in plasma containment devices so
using \eqref{eq:ucrosscond} to make a decomposition of $\vec{B}$
in terms of $\vec{u}$ analogous to the reverse in \eqref{eq:QEXB}
seems less useful.}

\subsection{The equilibrium ergodicity problem\label{subsec:EqmPhi}}

Resolving the IOL onto the vectors $\vec{B}$ and $\vec{v}$ (or $\vec{\ensuremath{u}}$)
eliminates the $\vec{v}\cross\vec{B}$ term, so in equilibrium, when
$\partial_{t}\vec{A}=0$ these components of the IOL imply

\begin{align}
\vec{B}\dotv\grad\Phi & =0\;,\label{eq:Bdotcond}\\
\vec{u}\dotv\grad\Phi & =0\;.\label{eq:udotcond}
\end{align}
This means $\Phi=\const$ on stream lines as well as magnetic field
lines. As a consequence, level sets of $\Phi$ are invariant under
magnetic and fluid flow. For instance, if $\Phi$ has smoothly nested
level surfaces in a region then both $\vec{u}$ and $\vec{B}$ lie
in the local tangent plane at each point on each isopotential surface
--- the magnetic and fluid flows are both locally integrable.

\begin{figure}
\centering \includegraphics[width=1\textwidth]{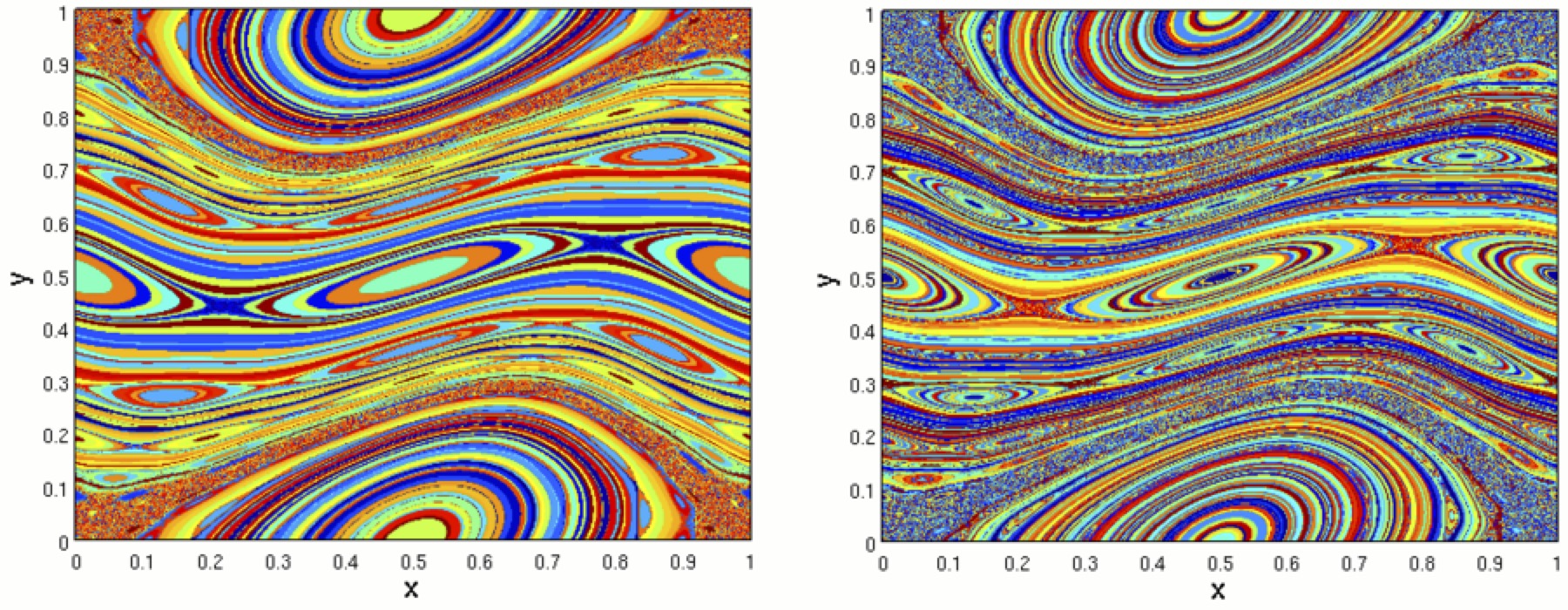}
\caption{Ergodic partition of iterates of the standard map as depicted in Fig.
11 of \citet{Levnajic_Mezic_2010}. (Reprinted with permission from
\emph{Chaos}.)\label{fig:ErgodPartition} }
\end{figure}

In the opposite extreme, \citet{Finn_Antonsen_83} {[}after Eq.~(29){]}
conclude from the constancy of $\Phi$ along a field line that ``if
the turbulent relaxation has ergodic field lines throughout the plasma
volume,'' then $\grad\Phi=0$, which implies that $\vec{u}\cross\vec{B}=0$
--- the fluid flows along magnetic field lines. As already mentioned,
we call such field-aligned steady flows \emph{fully relaxed equilibria}
(though the converse does not apply --- field-aligned flows can be
integrable).

However, field-aligned flow equilibria exclude many applications of
physical interest --- in particular tokamaks with strong toroidal
flow. For such axisymmetric equilibria \citet{Dewar_Burby_Qu_Sato_Hole_2020}
show RxMHD can give the same axisymmetric relaxed solutions \emph{with}
cross-field flow as found by \citet{Finn_Antonsen_83} and \citet{Hameiri_83},
but without needing the angular momentum constraint used by these
authors.

Unlike \citet{Finn_Antonsen_83} we are not appealing to turbulence
to justify relaxation, but, in fully three dimensional (3-D) plasmas,
we may be able to appeal to the existence of chaotic magnetic field
and stream lines. However ``chaotic'' is not the same as ``ergodic''
--- while chaotic flows do involve ergodicity, this is in an infinitely
complicated way, visualized in Figure~\ref{fig:ErgodPartition}{]}
in terms of the fractal ergodic partition of {[}\citet{Mezic_Wiggins_99,Levnajic_Mezic_2010}.
(This figure is generated for an iterated area-preserving map, but
magnetic field-line flows being flux preserving, the magnetic field-line
return map of a Poincaré section onto itself in a magnetic containment
device is similar.)

A similar problem involving chaos and ergodicity arises in magnetohydrostatics,
\citet{Hudson_etal_2012b}, where the equilibrium condition $\grad p=\vec{j}\cross\vec{B}$
implies $\vec{B}\dotv\grad p=0$, analogously to \eqref{eq:Bdotcond}
for $\Phi$, so the fractal ergodic partition for field-line flow
is as relevant to the pressure $p$ as it is for the potential $\Phi$.
In their MRxMHD equilibrium code \citet{Hudson_etal_2012b} solved
the puzzle posed by \citet{Grad_67} (i.e. how to formulate the three-dimensional
IMHD equilibrium problem so as to avoid a ``pathological'' pressure
profile) by using a much simpler ergodic partition obtained by aggregating
contiguous elements of the fractal ergodic partition into a finite
number of constant-pressure ``relaxation regions'' $\Omega_{i}$,
with pressure changing (discontinuously) only across the interfaces
between the $\Omega_{i}$s. The code was thus named the Stepped Presssure
Equilibrium Code (SPEC).

\subsubsection{Continuity of Electrostatic Potential\label{subsec:PhiCont}}

One might think that an analogous ``stepped potential equilibrium''
could provide a solution to the problem of finding a non-trivial but
tractable solution of $\grad_{\parallel}\Phi\defeq(\vec{B}/B)\dotv\grad\Phi=0$
in a chaotic magnetic-field-line flow. Unfortunately however we must
restrict $\grad_{\perp}\Phi\defeq(\Ident-\vec{B}\vec{B}/B^{2})\dotv\grad\Phi$
to square-integrable functions in order to keep the $\vec{E}\cross\vec{B}$
drift \eqref{eq:QEXB} from acquiring a $\delta$-function component.

This rules out having steps in $\Phi$ because $\delta$-functions
are not square integrable, so stepped potentials would make the kinetic
energy integral \emph{infinite}. However, this does not necessarily
imply $\Phi$ is constant in weakly chaotic regions with a finite
measure of KAM surfaces --- perhaps weak KAM theory {[}see e.g. \citet{Fathi_2009}{]}
would allow fractal potential profiles having finite kinetic energy
associated with them.

As a way to handle non-constant $\Phi$ computationally we propose
using a penalty or augmented Lagrangian method {[}see e.g. \citet{Nocedal_Wright_2006}{]}.
That is, we treat Hamilton's Principle as a constrained saddle-point
optimization problem and add a penalty functional to the Hamiltonian,
which regularizes the variational problem by approaching the (perhaps
fractal) IMHD ``feasible region'' of configuration space from outside,
in the less-constrained space on which RxMHD is defined {[}which is
smoother, see Figure 1 of \citet{Dewar_Burby_Qu_Sato_Hole_2020}{]}.

Another approach might be a time-evolution code with added dissipation
such that the long-time solution is attracted to one having chaotic
regions of constant pressure interspersed with integrable regions
with changing pressure. This can be viewed as a steepest-descents
solution of the same optimization problem.

\section{Constraints and Constrained Optimization\label{sec:Constraints} }

In this section we first discuss the new aspect of variational relaxation
theory introduced in this paper, namely the imposition of Ideal Ohm's
Law (IOL) as a constraint.

We then review use the Lagrange multiplier method in Subsection~\ref{subsec:macroPSL}
for imposing the conservation of entropy, magnetic helicity and cross
helicity as hard constraints, causing the EL equations, and hence
the conserved quantities, to be parametrized by the triplet of multipliers
$\tau_{\Omega}$, $\muOm$, and $\nuOm$ (the subscripts $\Omega$
indicating they are constant throughout $\Omega$, but may jump across
$\partial\Omega$ if there are adjacent relaxation regions as in MRxMHD).

\subsection{Augmented Lagrangian constraint method\label{subsec:AugLag}}

In implementing the IOL constraint we propose to adapt the \emph{Augmented
Lagrangian method} from finite-dimensional optimization theory, as
described by \citet[ §17.3]{Nocedal_Wright_2006}, or for Banach
spaces {[}see e.g. \citet{Kanzow_Steck_Wachsmuth_2018} and references
therein{]}. This is a hybrid numerical method that combines two constraint
approaches: the \emph{Lagrange multiplier method} and the \emph{penalty
function method}. We shall use the Lagrange multiplier method in Subsection~\ref{subsec:macroPSL}
for imposing the conservation of entropy, magnetic helicity and cross
helicity as hard constraints, causing the EL equations, and hence
the conserved quantities, to be parametrized by the triplet of multipliers
$\tau_{\Omega}$, $\muOm$, and $\nuOm$ (the subscripts $\Omega$
indicating they are constant throughout $\Omega$, but may jump across
$\partial\Omega$ if there are adjacent relaxation regions as in MRxMHD).

To impose the IOL as a hard constraint using the Lagrange multiplier
method we would ``simply'' add $\lamIOL\dotv\left(\vec{E}+\vec{v}\cross\vec{B}\right)$
to the Lagrangian density, solve the resultant EL equations to give
$\vec{E}+\vec{v}\cross\vec{B}$ as a function of the Lagrange multiplier
$\lamIOL$, and then solve for $\lamIOL$ such that $\vec{E}+\vec{v}\cross\vec{B}=0$.

Apart from the unavoidable complication that $\lamIOL$ is not just
a 3-vector but also is a function of $\vec{x}$ and $t$, so infinite
dimensional, there is the more fundamental problem, flagged in Subsection~\eqref{subsec:EqmPhi},
that the limit $\vec{E}+\vec{v}\cross\vec{B}\to0$ is likely singular
in 3-D equilibria because $\vec{E}$, $\vec{v}$, and $\vec{B}$ presumably
tend toward being fractal functions. Thus there is good reason to
believe the hard IOL constraint problem is ill-posed in 3-D systems
such as stellarators, which leads us to seek a soft IOL constraint
approach in order to regularize the Hamilton's Principle optimization
problem.

We build in the Maxwell-Faraday induction constraint \eqref{eq:MaxFad}
as a hard constraint by using the potential representations \eqref{eq:Erep},
$\vec{E}=-\grad\Phi-\epsF\partial_{t}\vec{A}$, and \eqref{eq:Brep},
$\vec{B}=\curl\vec{A}$. Thus the set of primary variables subject
to variation during an optimization is

\begin{equation}
X=\left\{ \vec{r},\vec{u},p,\vec{A},\Phi\right\} \,,\label{eq:stateVector}
\end{equation}
where $\vec{\ensuremath{r}}$ is the Lagrangian fluid-element position
field discussed in Section~\ref{sec:IMHDPSL}. {[}Note we have not
included $\rho$ and $\vec{v}$ as a independent variables because
they are functionals of $\vec{r}$, with variations given by \eqref{eq:rhovar}
and \eqref{eq:vvar}.{]}

The simplest soft IOL constraint approach is to add $\half\muIOL\left(\vec{E}+\vec{v}\cross\vec{B}\right)^{2}$
to the Hamiltonian density (thus subtracting it from our Lagrangian
density), where $\muIOL\to+\infty$ is a penalty multiplier. In this
limit the penalty term is supposed to dominate all other terms in
the Hamiltonian or Lagrangian and enforce IOL feasibility through
a sequence of infeasible solutions. However, this method is clearly
ill-conditioned numerically, leading us to resort to the ``best of
both worlds'' augmented Lagrangian method described below.

\subsubsection{The IOL as a softly constrained optimization problem\label{subsec:SoftIOL}}

In implementing the parallel IOL constraint we propose to adapt the
\emph{Augmented Lagrangian method} from finite-dimensional optimization
theory, as described by \citet[ §17.3]{Nocedal_Wright_2006}, or
for Banach spaces {[}see e.g. \citet{Kanzow_Steck_Wachsmuth_2018}
and references therein{]}. This is a hybrid numerical method that
combines two constraint approaches: the \emph{Lagrange multiplier
method} and the \emph{penalty function method} sketched above.

As well as adapting notation and methods from optimization theory
we have borrowed the terms \emph{feasible region}, meaning the range
in which the vector $X$ of variables to be solved for is such that
a set of \emph{equality} \emph{constraints} \textbf{$c_{i}[X]=0$
}are satisfied {[}also\emph{ inequality constraints} \textbf{$c_{i}[X]>0$},
but we do not consider this case{]}. The \emph{infeasible region},
is its complement, where one or more constraints are violated. By
\emph{hard constraint} we mean one where $X$ \emph{must} be in the
feasible region, and by \emph{soft constraint }we mean one where $X$
need only be in some \emph{neighbourhood }of the feasible region,
which is useful both practically and for regularizing when, as in
MHD, defining the boundary between feasible and infeasible is complicated
by the possibility of singular behaviour like current sheets and reconnection
points.

We now formulate two related physical tasks, the simpler one being

1. \textbf{The equilibrium problem:} In toroidal plasma confinement
theory the most physically desirable states are stable, time-independent
\emph{equilibria}, i.e. \emph{minima} of a Hamiltonian functional
$\mathscr{H}[X]$, kinetic \emph{plus} potential energy within a static
boundary $\partial\Omega$. We seek a numerical algorithm that starts
from an initial guess for the physical fields $X$ and iterates to
extremize (minimize if seeking a stable equilibrium) a Hamiltonian,
under the IOL equality constraint, \eqref{eq:idealOhm}. Finding a
stable equilibrium can be stated concisely as the optimization problem

\begin{equation}
\textbf{Equilibrium}\quad\min_{X}\mathscr{H}[X]\quad\text{subject to }\:\cIOL[X](\vec{x})]=0,\:\forall\,\vec{x}\in\Omega\;\text{and b.c.s\;}\:\forall\,\vec{x}\in\partial\Omega\;,\label{eq:EqProblem}
\end{equation}
where

\begin{equation}
\cIOL\defeq\vec{E}+\vec{v}\cross\vec{B}\;.\label{eq:cdef}
\end{equation}
is the generalization of \citet{Nocedal_Wright_2006}'s finite set
of equality constraint functions $\left\{ c_{i}\right\} $ (as a 3-vector
it is finite-dimensional but as a function of $\vec{x}$ it is infinite
dimensional).

For the purposes of the present paper $\mathscr{H}$ is the noncanonical
version, $H_{{\rm nc}}^{{\rm MHD}}$, of the Hamiltonian, $H_{{\rm }}^{{\rm MHD}}$
defined in \eqref{eq:Hgen} plus the global constraint terms described
in the next subsection, \ref{subsec:Invariants}. The ideal boundary
conditions (b.c.s) are $\vec{E}+\vec{v}\cross\vec{B}=0$, $\vec{\ensuremath{n}}\dotv\vec{v}=0$,
$\vec{\ensuremath{n}}\dotv\vec{B}=0$ on $\partial\Omega$ and $\Phi=\const$
on each disjoint component of $\partial\Omega$ (think plates of a
capacitor or electrodes of a vacuum tube).

To treat the implementation of the IOL in Hamilton's Principle, a
constrained saddle point optimization problem, we shall use the set
of values of the components of $\Ees$($\vec{\ensuremath{x}}\in\Omega$)
Note the identities 
\begin{equation}
\frac{\partial\cIOL}{\partial\vec{E}}=\Ident\;,\quad\frac{\partial\cIOL}{\partial\vec{v}}=\Ident\cross\vec{B}\quad\text{and}\quad\frac{\partial\cIOL}{\partial\vec{B}}=-\Ident\cross\vec{v}\;.\label{eq:cderivs}
\end{equation}

We seek a soft form of the equilibrium constraint, i.e. a formulation
such that $\cIOL\to0,\,\forall\,\vec{x}\in\Omega$, where $\to$ denotes
a limiting process whereby $X$ moves from the \emph{infeasible} class
of states where\textbf{ $\cIOL\neq0$} toward the \emph{feasible}
class defined pointwise as \textbf{$\cIOL=0,\,\forall\,\vec{x}\in\Omega(t)$},
or, in a weak form, as $||\cIOL||=0$. Such a soft constraint procedure
is provided by the \emph{augmented Lagrangian} (or, rather, Hamiltonian
in the Equilibrium problem) as prescribed by \citet[ §17.3]{Nocedal_Wright_2006},

\begin{equation}
\mathscr{H}_{{\rm A}}\defeq\int_{\Omega}\left[\mathcal{H}-\lamIOL\dotv\cIOL+\frac{1}{2}\muIOL\cIOL^{2}\right]\,\d V\;,\label{eq:AugmentedH}
\end{equation}
where $\lamIOL$ is a Lagrange multiplier and the spatial constant
$\muIOL\geq0$ is a \emph{penalty multiplier} of the non-negative
\emph{quadratic penalty} $\half\int_{\Omega}\cIOL^{2}\,\d V$.

\citet[ §17.3]{Nocedal_Wright_2006} give an iterative algorithmic
framework that combines the advantages of both the Lagrange multiplier
and penalty function methods. In their algorithm the user provides
an increasing sequence $\left\{ \muIOL|^{n},:n=0,1,2,\ldots\right\} $
penalty multipliers and adjusts $\lamIOL$ to solve for the minima
of Hamiltonians with the Laggrange muliplier and penalty terms. The
iteration update rule for the sequence of Lagrange multipliers and
corresponding constraint residuals $\left\{ \lamIOL^{n},\cIOL^{n}:n=0,1,2,\ldots\right\} $
is

\begin{align}
\text{\ensuremath{\lamIOL^{n+1}}} & =\lamIOL^{n}-\muIOL|^{n}\cIOL^{n}\;.\label{eq:lamUpdate}
\end{align}

In the following sections we shall take the iteration index $n$ as
implicit unless needed for clarity, with the updated $\lamIOL$ as
given by the RHS of \eqref{eq:lamUpdate} denoted by

\begin{equation}
\lamstar\defeq\frac{\partial}{\partial\cIOL}\left(\lamIOL\dotv\cIOL-\frac{\mu_{\Omega}^{{\rm P}}}{2}\cIOL^{2}\right)=\lamIOL-\mu_{\Omega}^{{\rm P}}\,\cIOL\;,\label{eq:lamstardef}
\end{equation}
wich is a ``best estimate'' of the optimum Lagrange multiplier given
the current estimate and penalty multiplier.

The more difficult second physical task is:

2.\textbf{ The time evolution problem:} This is similar to Task 1
except we seek an \emph{evolution, }a dynamical path in space-time
$\Omega_{t}\times[t_{1},t_{2}]$ given a time-dependent boundary $\partial\Omega_{t}$,
the objective function for extremization now being an action integral.
The stable minima of the Hamiltonian now becoming \emph{saddle points}
of the coresponding action functional $\mathscr{S}[X]$, kinetic \emph{minus}
potential energy. This task can be summarized as the pseudo optimization
problem 
\begin{equation}
\textbf{Dynamics}\quad\underset{X}{\text{extr}}\,\mathscr{S}[X]\quad\text{subject to}\;\cIOL\left[X\right](\vec{x},t)=0,\:\:\forall\,\vec{x}\in\Omega_{t},\,t\in[t_{1},t_{2}]\;,\label{eq:DynProblem-1}
\end{equation}
under the same boundary conditions as for equilibrium at each time
$t$.

Although sometimes called the ``Principle of Least Action'', Hamilton's
Principle is often \emph{not} an optimization problem but rather a
saddlepoint problem, where the stationary point of $\mathscr{S}[X]$
cannot be found by a descent algorithm. {[}This is well known in nonlinear
Hamiltonian dynamics\emph{ }{[}\citet{Meiss_92}{]} where periodic
orbits are classified as (action) \emph{minimizing} orbits, which
are hyperbolic (unstable), or as \emph{minimax} orbits, which are
elliptic (stable).{]} Although ``extremum'' or ``extremization''
is not quite correct either, as extremum strictly means ``maximum
\emph{or} mininimum'', it is convenient to use the abbreviation ``extr''
as an abbeviation for these words and add the rider ``depending on
direction of traversal'' (implying also the existence of neutral
directions between max and min), so as to include saddle points.

To find saddle points requires some form of Newton method, needing
at least estimates of the second variation (Hessian matrix) rather
than a descent method. The augmented Lagrangian method still works
if we solve \eqref{eq:DynProblem-1} at each iteration, so here again
we adopt it to solve for a stationary point of the augmented \emph{phase
spoce action functional }\eqref{eq:PSLnc} 
\begin{equation}
\mathscr{S_{{\rm ph}}^{{\rm {\rm A}}}}\defeq\iint_{\Omega}\!\left(\rho\vec{u}\dotv\vec{v}-\mathcal{H}+\LdenEXB\right)\d V\d t,\label{eq:AugmentedS}
\end{equation}
where the \emph{augmented penalty} \emph{constraint density} $\LdenEXB$
is defined by 
\begin{equation}
\LdenEXB\defeq\lamIOL\dotv\cIOL-\frac{\muIOL\,\cIOL^{2}}{2},\label{eq:ConstraintLden}
\end{equation}
with $\lamIOL$ and $\muIOL$ are taken as external parameters in
the application of Hamilton's Principle at each iteration, giving
a sequence of regularized magnetofluid models. When $\muIOL=0$, the
pure Lagrange multiplier method, feasible critical points of $\mathscr{H}_{{\rm A}}$
might be saddle points with descending directions in the infeasible
sector even if they are physically stable ideal equilibria where the
IMHD Hamiltonian is minimized. When $\lamIOL=0$, the pure penalty
function method, feasible stable equilibria could be approximated
arbitrarily well in the limit as $\muIOL$ tends to infinity, but
this becomes an increasingly ill-posed optimization problem. (It does
however have the attractive feature of providing a continous family
of relaxed MHD models running from the RxMHD of \citet{Dewar_Burby_Qu_Sato_Hole_2020}
when $\muIOL=0$ to a subset of weak IMHD when $\muIOL\to+\infty$.)

Remarks:

(i) ~Task 1 can be treated as a subclass of Task 2 in which time
derivatives are set to zero and $t$ is taken to be an irrelevant
constant, but the Hamiltonian is more appropriate than the Lagrangian
for treating it as an optimization problem.

(ii) The iteration method for implementing constraints is \emph{implicit},
meaning that the state variables in the \emph{$n^{{\rm th}}$ }iteration
need to be found by solving Euler--Lagrange equations, taking it
for granted the Euler--Lagrange equations can be solved and any sub-iterations
required have converged.

We shall not discuss detailed implementation issues here, except to
remark that time evolution over a large time interval can be implemented
numerically in an outer time-stepping loop in which a large time interval
is split into multiple short time intervals (timesteps) \textbf{$[t_{i},t_{i+1}]$},
within each of which constraint iterations are repeated until converged
to the required accuracy. Thus the evolutions required in implementing
the constraint iterations are over short time intervals, with each
initial guess being the converged evolution from the previous timestep
and the evolution representable to sufficent accuracy on a low-dimensional
interpolation basis (e.g. dimension 2 for piecewise-linear representation
of the full evolution) --- the increase in difficulty in going from
Task 1 to Task 2 may not be as great as at first it appears to be.

\subsection{Global constraints for isothermal RxMHD and IMHD\label{subsec:Invariants}}

We shall always retain the microscopic holonomic constraints (Section
\ref{sec:IMHDPSL}) on $\rho$ and $\vec{v}$, but we \emph{relax}
the infinite number of microscopic dynamical constraints on $p$ and
$\vec{B}$ imposed in IMHD by replacing these constraints with only
three \emph{macroscopic} hard constraints. These three constraints,
described below, are chosen to be quantities that are exact invariants
under IMHD dynamics in order to ensure that relaxed equilibria are
subset of all ideal equilibria. Further, as we seek plasma relaxation
formalisms applicable in arbitrary 3-D toroidal geometries, we invoke
only the MHD invariants least dependent on integrability of the fluid
and magnetic field line flows, the conservation of total mass $M_{\Omega}\defeq\int_{\Omega}\rho\,\d V$
being the most fundamental (whose conservation is built in microscopically)
. While these global invariants are not as well conserved as mass
under small resistive, viscous and 3-D chaos efects, in the spirit
of \citet{Taylor_86} we assume they are sufficiently robust that
postulating their conservation produces a model that is useful in
appropriate applications.

We can get IMHD by retaining \emph{all} the microscopic holonomic
constraints of Section~\ref{sec:IMHDPSL}, but it seems more physically
relevant to almost collisionless hot plasmas with high thermal conductivity
along magnetic field lines to relax the plasma \emph{thermally} by
relaxing the microscopic dynamical constraint on $p$ and replacing
it with the first global constraint below (entropy) to give\emph{
isothermal IMHD}.

As just indicated, our first global constraint is the adiabatic-ideal-gas
thermodynamic invariant, \emph{total entropy} 
\begin{equation}
S_{\Omega}[\rho,p]\defeq\int_{\Omega}\frac{\rho}{\gamma-1}\ln\left(\kappa\frac{p}{\rho^{\gamma}}\right)\,\d V\;,\label{eq:Entropy}
\end{equation}
where $\kappa$ is, for our purposes, an arbitrary dimensionalizing
constant, though it can be identified physically through a statistical
mechanical derivation of \eqref{eq:Entropy} {[}see e.g. \citet{Dewar_Yoshida_Bhattacharjee_Hudson_2015}{]}.
Its functional derivatives are 
\begin{align}
\frac{\delta S_{\Omega}}{\delta\rho} & =\frac{1}{\gamma-1}\ln\left(\kappa\frac{p}{\rho^{\gamma}}\right)-\frac{\gamma}{\gamma-1}\;,\label{eq:Entropydrho}\\
\frac{\delta S_{\Omega}}{\delta p} & =\frac{1}{\gamma-1}\,\frac{\rho}{p}\;.\label{eq:Entropydp}
\end{align}

We also impose conservation of the \emph{magnetic helicity} $2\muSI K_{\Omega}$,
where, following \citet{Bhattacharjee_Dewar_82}, we define the invariant
$K_{\Omega}$ as 
\begin{equation}
K_{\Omega}[\vec{A}]\defeq\frac{1}{2\muSI}\int_{\Omega}\vec{A}\dotv\vec{B}\,\d V\label{eq:Helicity}
\end{equation}
giving, with help of \eqref{eq:LemdFdA}, the functional derivative
\begin{equation}
\frac{\delta K_{\Omega}}{\delta\vec{A}}=\frac{\vec{B}}{\muSI}\;.\label{eq:HelicityFD}
\end{equation}

As discussed by \citet{Hameiri_2014}, in single-fluid IMHD we do
not have a separate fluid helicity invariant, but do have the \emph{cross
helicity} $\muSI K_{\Omega}^{{\rm X}}$, which can be derived from
a relabelling symmetry in the Lagrangian representation of the fields,
see e.g. Ch. 7 of \citet{Webb_2018}. Analogously to our other constraint
parameters containing $\vec{B}$, we include $\muSI^{-1}$ in the
definition of the cross helicity functional, 
\begin{equation}
K_{\Omega}^{{\rm X}}[\vec{u},\vec{A}]\defeq\frac{1}{\muSI}\int_{\Omega}\!\vec{u}\dotv\vec{B}\,\d V\;,\label{eq:XHel}
\end{equation}
which, like $P_{\Omega}$ and $S_{\Omega}$, has two functional derivatives
\begin{equation}
\frac{\delta K_{\Omega}^{{\rm X}}}{\delta\vec{u}}=\frac{\vec{B}}{\muSI}\,,\qquad\frac{\delta K_{\Omega}^{{\rm X}}}{\delta\vec{B}}=\frac{\vec{u}}{\muSI}\;.\label{eq:XHelFDs}
\end{equation}

\subsection{IOL-constrained Phase-Space Lagrangians and Actions\label{subsec:macroPSL}}

As foreshadowed, our recipe for constructing a non-dissipative relaxed
magnetofluid model is to start with the IMHD noncanonical Hamiltonian,
\eqref{eq:HMHD}, but to relax many, but not all, of the microscopic
constraints to which it is subject when deriving the IMHD Euler--Lagrange
equations. Specifically, to retain the basic compressible Euler-fluid
backbone of our relaxed MHD model \citet{Dewar_Yoshida_Bhattacharjee_Hudson_2015}
we keep the microscopic kinematic and mass conservation constraints,
\ref{eq:vvar} and \eqref{eq:rhovar}.

However we delete the microscopic ideal gas and flux-frozen magnetic
field variational constraints, $\delta p=-\gamma p\divv\bm{\xi}-\bm{\xi}\dotv\grad p$
and $\delta\vec{B}=\curl(\bm{\xi}\cross\vec{B})$, replacing these
infinities of constraints with only the three robust IMHD global invariants
(\ref{eq:Entropy}--\ref{eq:XHel}). These global constraints are
imposed by adding the \emph{global-invariants-constraint} (GIC) Lagrange
multiplier term

\begin{equation}
\mathcal{L}_{{\rm \Omega}}^{{\rm GIC}}\defeq\tau_{\Omega}\frac{\rho}{\gamma-1}\ln\left(\kappa\frac{p}{\rho^{\gamma}}\right)+\muOm\frac{\vec{A}\dotv\vec{B}}{2\muSI}+\nuOm\frac{\vec{u}\dotv\vec{B}}{\muSI}\label{eq:GIcstr}
\end{equation}
to $\mathcal{L}_{{\rm nc}}^{{\rm MHD}}$ to form the \emph{RxMHD PSL
density} {[}\citet{Dewar_Burby_Qu_Sato_Hole_2020}{]}

\begin{align}
\LdenRx & \defeq\mathcal{L}_{{\rm nc}}^{{\rm MHD}}+\mathcal{L}_{{\rm \Omega}}^{{\rm GIC}}=\rho\vec{u}\dotv\vec{v}-\frac{\rho\vec{u}^{2}}{2}-\frac{p}{\gamma-1}-\frac{B^{2}}{2\muSI}\nonumber \\
 & \qquad\qquad\qquad\qquad\quad+\tau_{\Omega}\frac{\rho}{\gamma-1}\ln\left(\kappa\frac{p}{\rho^{\gamma}}\right)+\muOm\frac{\vec{A}\dotv\vec{B}}{2\muSI}+\nuOm\frac{\vec{u}\dotv\vec{B}}{\muSI}\;.\label{eq:RxPSL}
\end{align}

In \ref{eq:GIcstr} the Lagrange multipliers $\tau_{\Omega},\muOm,$
and $\nuOm$ are spatially constant throughout $\Omega$, but can
change in time to enforce constancy respectively of total entropy,
magnetic helicity and cross helicity in $\Omega$. By removing the
infinite numbers of microscopic constraints on $p$ and $\vec{B}$
that are imposed in IMHD, in the RxMHD formalism \citet{Dewar_Burby_Qu_Sato_Hole_2020}
we greatly increased the variationally feasible region of the state
space, thus allowing the system to access a lower energy equilibrium.
In fact, as the Eulerian fields $\delta p(\vec{x},t)$ and $\delta\vec{A}(\vec{x},t)$
are now locally free variations at each point $\vec{x}$, we have
added two infinities of degrees of freedom, which turns out to be
too many as the IOL constraint embedded in IMHD is entirely lost in
RxMHD.

Thus we reduce the degrees of freedom of RxMHD by imposing a soft
penalty-function IOL constraint using the augmented Lagrangian constraint
density $\LdenEXB$, \eqref{eq:ConstraintLden}. As the IOL constraint
applies pointwise throughout $\Omega$, giving an infinite number
of constraints, on $\Phi$ and $\vec{B}$. Adding the constraint term
we get the full Lagrangian density with augmented constraint

\begin{equation}
\LdenQRx\defeq\LdenRx+\LdenEXB\;.\label{eq:QRxPSL}
\end{equation}

We shall also have need to define the \emph{gauge-invariant part}
of the Lagrangian density by substracting off the magnetic helicity
term,

\begin{align}
\LdenQPr & \defeq\LdenQRx-\muOm\frac{\vec{A}\dotv\vec{B}}{2\muSI}\;.\label{eq:LAminusdef}
\end{align}
(For derivatives of the Lagrangian density with respect to anything
other than $\vec{{A}}$, $\vec{{B}}$, $\vec{x}$, or $t$, $\LdenQRx$
and $\LdenQPr$ can be used interchangeably.)

The augmented phase-space action integral is 
\begin{equation}
\SQRx=\int\!\!\d t\!\int_{\Omega}\!\!\d V\,\LdenQRx\;.\label{eq:SQRx}
\end{equation}
As in \eqref{eq:PSLnc}, the fluid velocity $\vec{u}$ is treated
as a noncanonical momentum variable that is freely variable in the
phase-space version of Hamilton's Principle, $\delta\SQRx=0$, and
$\vec{v}$ is a relative flow whose variation with respect to $\bm{\xi}$
obeys the kinematical constraint \eqref{eq:vvar}. It is also the
flow appearing in the mass conservation constraint equations \eqref{eq:rhovar}
and \eqref{eq:Continuity}.

\section{Euler--Lagrange (EL) equations\label{sec:RxMHDEL}}

\subsection{Formal view of EL equations\label{subsec:QRxMHDSvar}}

The utility of Hamilton's action-principle approach is that a complete
set of equations for our physical variables is provided by the EL
equations following from the general variation of the generic augmented
action $\SQRx$, 
\begin{align}
\delta\SQRx & =\int\!\!\d t\!\!\int_{\Omega}\!\!\!\d V\,\!\!\left[\delta\vec{u}\dotv\frac{\partial\LdenQRx}{\partial\vec{u}}+\delta\vec{A}\dotv\frac{\partial\LdenQRx}{\partial\vec{A}}+\delta\vec{B}\dotv\frac{\partial\LdenQRx}{\partial\vec{B}}+\delta\vec{E}\dotv\frac{\partial\LdenQRx}{\partial\vec{E}}\right.\nonumber \\
 & \left.\qquad\qquad\qquad\qquad\qquad\quad+\,\delta p\frac{\partial\LdenQRx}{\partial p}+\delta\rho\frac{\partial\LdenQRx}{\partial\rho}+\delta\vec{v}\dotv\frac{\partial\LdenQRx}{\partial\vec{v}}\right]\nonumber \\
 & \defeq\int\!\!\d t\!\!\int_{\Omega}\!\!\!\d V\,\!\!\left[\delta\vec{u}\dotv\frac{\delta\SQRx}{\delta\vec{u}}+\delta\Phi\frac{\delta\SQRx}{\delta\Phi}+\delta\vec{A}\dotv\frac{\delta\SQRx}{\delta\vec{A}}+\delta p\frac{\delta\SQRx}{\delta p}+\bm{\xi}\dotv\frac{\delta\SQRx}{\delta\vec{x}}\right]\label{eq:genVar}
\end{align}
where the top equation on the RHS is simply an integral over the first
variation of $\LdenQRx$ and the second RHS equation\emph{ defines}
the functional derivatives with respect to the independent variables
by matching the corresponding terms in the top RHS equation after
the variations of the explicit variables in $\LdenQRx$ are expanded
and integrations by parts where necessary --- ignoring boundary terms
as we can assume the support of the variations does not include the
boundary {[}note that there are no $\delta\lambda$ or $\delta\muIOL$
terms as $\lambda$ and $\muIOL$ are taken as given --- see discussion
around \eqref{eq:ConstraintLden}{]}.For instance $\delta\SQRx/\delta\vec{x}$
is the sum of the terms linear in $\bm{\xi}$ obtained from $\delta\rho$
and $\delta\vec{v}$ given in \eqref{eq:rhovar}, and \eqref{eq:vvar}respectively.
(Note: For notational convenience $\delta\vec{x}$ is used in the
denominator of the functional derivative as an alternative to the
Lagrangian variation of $\vec{x}$, denoted everywhere else as $\Delta\vec{x}$
or $\bm{\xi}$. It does \emph{not} denote the Eulerian variation of
$\vec{x}$, which is by definition zero.)

Inspecting \eqref{eq:ConstraintLden} we see that $\LdenEXB$ contains
$\vec{E}$ and $\vec{\ensuremath{B}}$ but does not contain $\vec{u}$,
$p$, or $\rho$, and no term in $\LdenRx$ contains $\vec{\ensuremath{E}}$,
$\grad\vec{u},\grad p,$ or $\grad\rho$, so the corresponding \emph{functional}
derivatives of $\SQRx$ are are simply \emph{partial} derivatives
of $\LdenQRx$, e.g. the $\delta\vec{u}$- and $\delta p$-EL equations
are 
\begin{equation}
\frac{\delta\SQRx}{\delta\vec{u}}=\frac{\partial\LdenQRx}{\partial\vec{u}}=0\;,\label{eq:deltauEL}
\end{equation}

\begin{equation}
\frac{\delta\SQRx}{\delta p}=\frac{\partial\LdenQRx}{\partial p}=0\text{\;}.\label{eq:deltapEL}
\end{equation}
The $\delta\vec{A}$-EL equation is best displayed by splitting $\LdenQRx$
into the gauge-invariant part $\LdenQPr$, \eqref{eq:LAminusdef},
and the magnetic helicity constraint term \textbf{$\muOm\vec{A}\dotv\vec{B}/2\muSI$}
in order to make manifest the explicit$\vec{A}$-dependence. Thus
\begin{equation}
\frac{\partial\LdenQRx}{\partial\vec{A}}=\frac{\muOm\vec{B}}{2\muSI}\;,\quad\frac{\partial\LdenQRx}{\partial\vec{B}}=\frac{\partial\LdenQPr}{\partial\vec{B}}+\frac{\muOm\vec{A}}{2\muSI}\label{eq:dLQdadB}
\end{equation}
The $\delta\vec{\!A}$-EL equation is then found by using these results
in the lemma \eqref{eq:LemdFdA} to give

\[
\frac{\delta\SQRx}{\delta\vec{A}}=\frac{\muOm\vec{B}}{2\muSI}+\curl\frac{\partial\LdenQPr}{\partial\vec{B}}+\curl\frac{\muOm\vec{A}}{2\muSI}+\epsF\frac{\partial}{\partial t}\frac{\partial\LdenQPr}{\partial\vec{E}}=0\;,
\]
i.e.

\begin{align}
\epsF\frac{\partial}{\partial t}\frac{\partial\LdenQPr}{\partial\vec{E}}+\curl\frac{\partial\LdenQPr}{\partial\vec{B}} & =-\frac{\muOm\vec{B}}{\muSI}\;.\label{eq:deltaAEL}
\end{align}
The $\delta\Phi$-EL equation is, using \eqref{eq:LemdFdPhi}, 
\begin{flushleft}
\begin{equation}
\frac{\delta\SQRx}{\delta\Phi}=\divv\frac{\partial\LdenQRx}{\partial\vec{E}}=0\;.\label{eq:deltaPhiEL}
\end{equation}
and, using \eqref{eq:vvar}, $\delta\vec{v}=\partial_{t}\bm{\xi}+\vec{v}\dotv\grad\bm{\xi}-\bm{\xi}\dotv\grad\vec{v}$,
the $\Delta\vec{x}$-EL {[}or $\bm{\xi}$-EL --- see \eqref{eq:genVar}{]}
equation is 
\begin{align}
\frac{\delta\SQRx}{\delta\vec{x}} & =-\partial_{t}\bm{\Pi}-\divv\left(\vec{v}\bm{\Pi}\right)-(\grad\vec{v})\dotv\bm{\Pi}\nonumber \\
 & \quad+\grad\left(\rho\frac{\partial\LdenQRx}{\partial\rho}\right)-(\grad\rho)\frac{\partial\LdenQRx}{\partial\rho}=0\;,\label{eq:xiEL}
\end{align}
where 
\begin{align}
\bm{\Pi} & \defeq\frac{\partial\LdenQRx}{\partial\vec{v}}\label{eq:PiDef}
\end{align}
is a new canonical momentum density (cf. $\bm{\pi}$ in Subsection
\eqref{eq:microPSL}). 
\par\end{flushleft}

\subsection{Formal conservation-form momentum equation\label{subsec:Conservation-form}}

A general form of the equation of motion is provided by the $\bm{\xi}$-EL
equation \eqref{eq:xiEL}, which agrees with (21) of \citet{Dewar_Burby_Qu_Sato_Hole_2020}
in the special case of their $\bm{\uplambda}=[\rho]$, $\bm{V}=[0]$,
and $\bm{\Lambda}=[1]$.

To get a more transparent version we now derive a canonical-momentum
conservation form of the equation of motion, the existence of which
is implied by Noether's theorem and translational invariance (\emph{within}
$\Omega$, i.e. not including\textbf{ $\partial\Omega$}). To do this
we transform \eqref{eq:xiEL} into the same form as (22) of \citet{Dewar_Burby_Qu_Sato_Hole_2020}\footnote{Unfortunately the seemingly general stress tensor (27) derived by
\citet{Dewar_Burby_Qu_Sato_Hole_2020} was limited to scalar fields
like $\Phi$. Appendix~\ref{sec:lemmas} derives \eqref{eq:Lem3}
to handle vector fields like $\vec{A}$.} by subtracting $\grad\LdenQRx\,$ from both sides, giving, after
a little rearrangement, 
\begin{align}
 & \partial_{t}\bm{\Pi}+\divv\left[\vec{v}\,\bm{\Pi}+\vsf{I}\left(\LdenQRx-\rho\frac{\partial\LdenQRx}{\partial\rho}\right)\right]\nonumber \\
 & =
\grad\LdenQRx-(\grad\vec{v})\dotv\bm{\Pi}-(\grad\rho)\,\frac{\partial\LdenQRx}{\partial\rho}\;,\label{eq:DeltaxELmod}
\end{align}

Local translational invariance implies the only $\vec{x}$ dependence
of $\LdenQRx$ is through its component fields, so the chain rule
gives

\begin{flalign*}
\grad\LdenQRx & =\left(\grad\vec{u}\right)\dotv\frac{\partial\LdenQRx}{\partial\vec{u}}+\left(\grad p\right)\frac{\partial\LdenQRx}{\partial p}\\
 & \quad+\left(\grad\vec{B}\right)\dotv\frac{\partial\LdenQRx}{\partial\vec{B}}+\left(\grad\vec{A}\right)\dotv\frac{\partial\LdenQRx}{\partial\vec{A}}+\left(\grad\vec{E}\right)\dotv\frac{\partial\LdenQRx}{\partial\vec{E}}\\
 & \quad+\left(\grad\rho\right)\,\frac{\partial\LdenQRx}{\partial\rho}+\left(\grad\vec{v}\right)\dotv\bm{\Pi}+\left(\grad\lamIOL\right)\dotv\frac{\partial\LdenQRx}{\partial\lamIOL}\;,
\end{flalign*}
which can be simplified slightly because the two terms on the top
line of the RHS vanish by \eqref{eq:deltauEL} and \eqref{eq:deltapEL}
. Using also \eqref{eq:dLQdadB} we get

\begin{flalign*}
\grad\LdenQRx & =(\grad\vec{B})\dotv\frac{\partial\LdenQPr}{\partial\vec{B}}+\frac{\muOm}{2\muSI}\left[(\grad\vec{B})\dotv\vec{A}+(\grad\vec{A})\dotv\vec{B}\right]+\left(\grad\vec{E}\right)\dotv\frac{\partial\LdenQRx}{\partial\vec{E}}\\
 & \quad+\left(\grad\rho\right)\,\frac{\partial\LdenQRx}{\partial\rho}+\left(\grad p\right)\,\frac{\partial\LdenQRx}{\partial p}+\left(\grad\vec{v}\right)\dotv\bm{\Pi}+\left(\grad\lamIOL\right)\dotv\frac{\partial\LdenQRx}{\partial\lamIOL}\\
 & =\left(-\epsF\frac{\partial}{\partial t}\frac{\partial\LdenQRx}{\partial\vec{E}}\right)\cross\vec{B}-\divv\left[\frac{\partial\LdenQPr}{\partial\vec{B}}\!\cross\Ident\cross\vec{B}\right]+\frac{\muOm}{2\muSI}\grad\left(\vec{A}\dotv\vec{B}\right)\\
 & \quad+\left(\grad\vec{E}\right)\dotv\frac{\partial\LdenQRx}{\partial\vec{E}}+\left(\grad\rho\right)\,\frac{\partial\LdenQRx}{\partial\rho}+\left(\grad\vec{v}\right)\dotv\bm{\Pi}+\left(\grad\lamIOL\right)\dotv\frac{\partial\LdenQRx}{\partial\lamIOL}\\
 & =\divv\left[\frac{\partial\LdenQPr}{\partial\vec{E}}\vec{E}-\frac{\partial\LdenQPr}{\partial\vec{B}}\!\cross\Ident\cross\vec{B}+\Ident\,\frac{\muOm\vec{A}\dotv\vec{B}}{2\muSI}\right]-\epsF\frac{\partial}{\partial t}\left(\frac{\partial\LdenQPr}{\partial\vec{E}}\cross\vec{B}\right)\\
 & \quad+\left(\grad\rho\right)\,\frac{\partial\LdenQPr}{\partial\rho}+\left(\grad\vec{v}\right)\dotv\bm{\Pi}+\left(\grad\lamIOL\right)\dotv\frac{\partial\LdenQPr}{\partial\lamIOL}\;,
\end{flalign*}
where we used the identity \eqref{eq:LemgradBdotf}, $(\grad\vec{B})\dotv\vec{f}=\left(\curl\vec{f}\right)\cross\vec{B}-\divv\left[\vec{f}\cross\Ident\cross\vec{B}\right]$
of Appendix\ \ref{sec:lemmas}, with $\vec{f}=\partial\LdenQPr/\partial\vec{B}$
and the $\delta\vec{\!A}$-EL equation \eqref{eq:deltaAEL}. Also
the identity \eqref{eq:LemgradEdotf} $\left(\grad\vec{E}\right)\dotv\vec{f}=\divv[\vec{f}\vec{E}]-\vec{E}\divv\vec{f}-\epsF\vec{f}\cross\partial_{t}\vec{B}$
with $\vec{f}=\partial\LdenQPr/\partial\vec{E}$, and the $\delta\Phi$-EL
equation \eqref{eq:deltaPhiEL}, to reduce all but the last three
terms to divergence form. Eliminating these $\grad\rho$ and $\grad\vec{v}$
terms between those in \eqref{eq:DeltaxELmod} and $\grad\LdenQRx$
above, and also cancelling the $\vec{A}\dotv\vec{B}$ terms, gives
a general momentum equation in gauge-independent conservation form
on the LHS, but with the $\grad\lamIOL$ term on the RHS acting as
an external forcing term,

\begin{equation}
\partial_{t}\left(\bm{\Pi}+\epsF\frac{\partial\LdenQPr}{\partial\vec{E}}\cross\vec{B}\right)+\divv\vsf T=\left(\grad\lamIOL\right)\dotv\frac{\partial\LdenQPr}{\partial\lamIOL}\;,\label{eq:MomConsvnForm}
\end{equation}
(where LHS/RHS denote ``left/right-hand side''). Here the tensor
$\vsf T$ is given by

\begin{align}
\vsf T & =\vec{v}\bm{\Pi}+\frac{\partial\LdenQPr}{\partial\vec{B}}\!\cross\Ident\cross\vec{B}-\frac{\partial\LdenQPr}{\partial\vec{E}}\vec{E}+\Ident\left(\LdenQPr-\rho\,\frac{\partial\LdenQPr}{\partial\rho}\right)\;.\label{eq:Tdef}
\end{align}
{[}See \eqref{eq:LemXIX} for a dyadic identity that is useful for
interpreting the second term on the RHS.{]}

This construction illustrates that the momentum conservation form
is a general property of any translation-invariant Lagrangian formulation
(by Noether's theorem) and thus is preserved even with our augmented
penalty function constraint (except for the forcing term from the
symmetry-breaking Lagrange multiplier). It is not manifestly symmetric
but we expect it to be symmetrizable from local rotational invariance
{[}\citet{Dewar_70}, \citet{Dewar_77}{]}.

We now examine the implications of our EL equations in more detail.

\subsection{Explicit Variation of Eulerian velocity}

From the $\delta\vec{u}$-EL equation \eqref{eq:deltauEL} , 
\begin{equation}
\rho(\vec{v}-\vec{u})+\frac{\nuOm\vec{B}}{\muSI}=0\;,\label{eq:deltauEL-1}
\end{equation}
which is equivalent to the relative flow formula \eqref{eq:vXshift}
given in the Introduction, thus both motivating and justifying substituting
$\vec{v}$ for $\vec{u}$ {[}see text below \eqref{eq:vXshift}{]}.
We shall use this below in the form $\vec{v}=\vec{u}-\uRx$ for eliminating
$\vec{v}$ when required. (Recall $\uRx\defeq\nuOm\vec{B}/\muSI\rho$.)

N.B. Taking the divergence of both sides of \ref{eq:vXshift}, the
EL equation \eqref{eq:deltauEL-1}, we have $\divv(\rho\vec{v})=\divv(\rho\vec{u})$.
Thus, as noted below \eqref{eq:vXshift}, $\vec{u}$ obeys the same
continuity equation as $\vec{v}$, \eqref{eq:Continuity}. That is,
\begin{equation}
\partial_{t}\rho+\divv(\rho\vec{u})=0\;.\label{eq:ucont}
\end{equation}

\subsection{Variation of pressure}

From the $\delta p$-EL equation \eqref{eq:deltapEL}

\[
\frac{1}{\gamma-1}\left(\frac{\tau_{\Omega}\rho}{p}-1\right)=0\;,
\]
which leads to the \emph{isothermal equation of state} 
\begin{equation}
p=\tau_{\Omega}\rho\;.\label{eq:deltapEL-1}
\end{equation}

A related result is sometimes useful: From \eqref{eq:QRxPSL} after
a little algebra, 
\begin{align}
\frac{\partial\LdenQPr}{\partial\rho} & =\vec{u}\dotv\vec{v}-\frac{\vec{u}^{2}}{2}-\tau_{\Omega}\ln\frac{\rho}{\rho_{\Omega}}\;,\nonumber \\
 & \defeq\vec{u}\dotv\vec{v}-h_{\Omega}\label{eq:dLRxdrho}
\end{align}
where the Bernoulli ``head'' $h_{\Omega}$ is defined by 
\begin{equation}
\begin{split}h_{\Omega} & =\frac{u^{2}}{2}+\tau_{\Omega}\ln\frac{\rho}{\rho_{\Omega}}\;,\end{split}
\label{eq:Head}
\end{equation}
with $\rho_{\Omega}$ a spatially constant reference density that
need not be given as it does not contribute to the $\Delta\vec{x}$-EL,
\eqref{eq:xiEL}. It has the property that $\rho\grad h_{\Omega}=\rho\grad\half u^{2}+\grad p$.

\subsection{Explicit Variation of scalar potential\label{subsec:PhiVariation}}

From the $\delta\Phi$-EL equation \eqref{eq:deltaPhiEL} and \eqref{eq:dLCdE},
\begin{align}
\divv\left(\frac{\partial\LdenEXB}{\partial\vec{E}}\right) & =\divv\left(\frac{\partial\LdenEXB}{\partial\vec{C}}\right)=0\nonumber \\
\text{i.e. }\divv\lamstar & =0\;,\label{eq:divlamstar}
\end{align}
where $\lamstar=\lamIOL-\muIOL\,\cIOL$ is as defined in \eqref{eq:lamstardef}.

Comparing the update rule \eqref{eq:lamUpdate}, $\text{\ensuremath{\lamIOL|^{n+1}}}=\lamIOL|^{n}-\muIOL|^{n}\vec{C}|^{n}$,
with \eqref{eq:lamstardef} we identify $\lamstar$ as the updated
$\lamIOL$ for initializing the next iteration, i.e. $\lamIOL|^{n+1}=\lamstar|^{n}$.
As $\divv\lamstar|^{n}=0$ we therefore have $\divv\lamIOL^{n+1}=0$,
and likewise for $\divv\lamIOL|^{n+2}$ and all subsequent Lagrange
multipliers in the iteration sequence. In fact, assuming integer $n$
is a typical step in the iteration, we must also conclude 
\begin{equation}
\divv\lamIOL|^{n}=0\:\;\forall\:n,\:\text{including}\:0\:\text{and }\infty\;.\label{eq:divlambda}
\end{equation}
Thus we can eliminate both $\lamIOL$ and $\lamstar$ from \eqref{eq:lamstardef}
by taking the divergence of both sides to give 
\begin{equation}
\divv\cIOL=0\;,\label{eq:divc}
\end{equation}
which, being a homogeneous equation, provides no driving term for
$\cIOL$.

While, from \eqref{eq:cdef}, \eqref{eq:divc} implies an inhomogeneous
equation for $\vec{E}$,

\begin{equation}
\divv\vec{E}=-\divv\left(\vec{u}\cross\vec{B}\right)\;,\label{eq:divE}
\end{equation}
this is also implied by the IOL, again showing that we cannot determine
non-feasibility by taking divergences only.

However, we also have an expression for $\curl\vec{E}$ from the Maxwell-Faraday
induction equation \eqref{eq:MaxFad}, which, combined with \eqref{eq:cdef},
gives the inhomogeneous equation

\begin{equation}
\curl\cIOL=-\epsF\partial_{t}\vec{B}+\curl\left(\vec{u}\cross\vec{B}\right)\;.\label{eq:curlc}
\end{equation}
Thus the non-feasibility parameter $\cIOL$ can be viewed as driven
by the departure from the ideal MHD magnetic-field evolution equation.

This is seen better by rewriting \eqref{eq:curlc} as an evolution
equation for $\vec{B}$,

\begin{equation}
\epsF\partial_{t}\vec{B}=\curl\left(\vec{u}\cross\vec{B}\right)-\curl\cIOL\;.\label{eq:curlc-1}
\end{equation}
When $\cIOL=0$ this is the IMHD evolution eqation for $\vec{B}$,
\emph{irrespective} of our magnetic and cross-helicity constraints
and confirms that allowing $\cIOL\neq0$ is sufficient to relax the
flux-freezing topological constraints of IMHD.

However, to satisfy \eqref{eq:MaxFad} automatically we use the potential
representations \eqref{eq:Erep}, $\vec{E}=-\grad\Phi-\epsF\partial_{t}\vec{A},$and
\eqref{eq:Brep}, $\vec{B}=\curl\vec{A}$, so \eqref{eq:cdef} becomes

\begin{equation}
\cIOL=-\grad\Phi-\epsF\partial_{t}\vec{A}+\vec{v}\cross\left(\curl\vec{A}\right)\;,\label{eq:cdefpot}
\end{equation}
showing $\cIOL$ as\emph{ the discrepancy between the potential representations
of} $-\vec{E}$ \emph{and of} $\vec{v}\cross\vec{B}$ (equivalently
$\vec{u}\cross\vec{B}$ after Euler--Lagrange equations are derived).
As \eqref{eq:cdefpot} implies \eqref{eq:curlc-1}, the latter is
now not an independent equation and is useful only for insight.

In potential representation \eqref{eq:divE} becomes the Poisson equation
\begin{align}
\nabla^{2}\Phi & =\divv\left(\vec{u}\cross\vec{B}\right)\;,\label{eq:PoissonEq}
\end{align}
where Coulomb gauge, $\divv\vec{A}=0$, has been adopted to eliminate
the explicit unknown $\vec{A}$, though it is still implicit through
$\vec{B}$. The solution of this elliptic differential equation, using
the Dirichlet boundary conditions discussed after \eqref{eq:cdef},
is such that $\Phi$ is a smooth function. However, as discussed in
Section~\ref{subsec:EqmPhi} the electric and magnetic field lines
it defines are not generically integrable and thus may represent chaotic
flows in 3-D geometries.

Assuming $\vec{u}$ and $\vec{A}$ are determined using other Euler--Lagrange
equations, so \eqref{eq:PoissonEq} can be solved for $\Phi$, \eqref{eq:cdefpot}
gives us $\cIOL$ and then $\lamstar$ from \eqref{eq:lamstardef}.
This will be illustrated in Subsection~\ref{subsec:WKB_IOLRxMHD}.

\subsection{Explicit Variation of vector potential \eqref{eq:QEXB}}

From \eqref{eq:LAminusdef}, \eqref{eq:RxPSL} and \eqref{eq:cderivs}
we have

\begin{align}
\frac{\partial\LdenQPr}{\partial\vec{B}} & =\frac{\partial}{\partial\vec{B}}\left(-\frac{B^{2}}{2\muSI}+\nuOm\frac{\vec{B}\dotv\vec{u}}{\muSI}+\cIOL\dotv\lamIOL-\frac{\muIOL\,\cIOL^{2}}{2}\right)\nonumber \\
 & =-\frac{\vec{B}}{\muSI}+\frac{\nuOm\,\vec{u}}{\muSI}+\frac{\partial\cIOL}{\partial\vec{B}}\dotv\lamstar\nonumber \\
 & =-\frac{\vec{B}}{\muSI}+\frac{\nuOm\,\vec{u}}{\muSI}-\vec{v}\cross\lamstar\label{eq:dLCdB}\\
\text{and}\quad\frac{\partial\LdenQPr}{\partial\vec{E}} & =\lamstar\;.\label{eq:dLCdE}
\end{align}
Inserting these identities in the $\delta\!\vec{A}$-EL equation \eqref{eq:deltaAEL}
gives

\begin{equation}
\epsF\frac{\partial\lamstar}{\partial t}-\curl\left(\vec{v}\cross\lamstar\right)=\frac{1}{\muSI}\left(\curl\vec{B}-\muOm\vec{B}-\nuOm\,\curl\vec{u}\right)\;,\label{eq:deltaAEL-1}
\end{equation}
displayed as an inhomogeneous hyperbolic equation for the Lagrange-multiplier
field $\lamstar$. However, it can also be displayed as an inhomogeneous
elliptic equation for $\vec{B}$ by multiplying both sides with $-\muSI$
and rearranging to give

\begin{align}
\curl\vec{B} & =\muOm\vec{B}+\nuOm\bm{\omega}+\muSI[\epsF\partial_{t}\lamstar-\curl\left(\vec{v}\cross\lamstar\right)]\;,\label{eq:ModBel}
\end{align}
where $\bm{\omega}\defeq\curl\vec{u}$ is the fluid \emph{vorticity}.
Apart from the terms in $\lamstar$ this is the RxMHD \emph{modified
Beltrami equation} found by \citet{Dewar_Burby_Qu_Sato_Hole_2020}.

The relation between $\Phi$, $\vec{A}$, $\lamstar$ and $\vec{C}$
appears somewhat difficult to untangle in general so we shall defer
detailed analysis of these equations to Section~\ref{sec:WKBRxMHD},
where the WKB aproximation makes the task easier. Suffice it here
simply to count equations to give confidence that the problem can
be solved in principle --- the four independent equations for these
four unknowns are, in order of occurrence, \eqref{eq:lamstardef},
\eqref{eq:cdefpot}, \eqref{eq:PoissonEq} and \eqref{eq:ModBel}.
{[}Unless we set $\nuOm=0$, $\rho$ occurs through the $\uRx$ in
$\vec{v}$, in which case we need to add \eqref{eq:ufullRx}, \eqref{eq:vXshift}
and \eqref{eq:ucont} to the list.{]} When solved, all variables should
be known in terms of $\vec{u}$, whose evolution can then be determined
from the $\bm{\xi}$-EL equation.

A final remark: Taking the divergence of both sides the $\delta\!\vec{A}$
Euler--Lagrange equation \eqref{eq:ModBel} verifies that it propagates
the $\delta\Phi$ Euler--Lagrange equation \eqref{eq:divlamstar},
$\divv\lamstar=0$. That is, if $\divv\lamstar=0$ initially, it will
remain so even if $\lamstar$ changes as the plasma evolves in time,
and at each step in the iteration to converge $\cIOL\to0$. So the
two Euler--Lagrange equations are consistent, though otherwise independent.

\subsubsection{Electric current}

We can also identify the \emph{electric current}, $\vec{j}\defeq\curl\vec{B}/\muSI$,
so \eqref{eq:ModBel} can be written 
\begin{align}
\vec{j} & =\frac{\muOm}{\muSI}\vec{B}+\frac{\nuOm}{\muSI}\bm{\omega}-\curl\left(\vec{v}\cross\lamstar\right)+\epsF\frac{\partial\lamstar}{\partial t}\nonumber \\
 & =\frac{\muOm}{\muSI}\vec{B}+\frac{\nuOm}{\muSI}\curl\left(\vec{u}+\frac{\vec{B}}{\rho}\cross\lamstar\right)-\curl\left(\vec{u}\cross\lamstar\right)+\epsF\frac{\partial\lamstar}{\partial t}\;.\label{eq:j}
\end{align}
The first term on the RHS of \eqref{eq:ModBel} is the usual parallel
electric current term of the linear-force-free (Beltrami) magnetic
field model, the second term is a \emph{vorticity-driven current}
\citet{Yokoi_2013} term, while the last term is a new \emph{IOL}
\emph{constraint current} which, (taking into account the EL equation
$\divv\lamstar=0$) maintains the divergence-free nature of $\vec{j}$
as required to maintain quasi-neutrality).

\subsubsection{Physical interpretation of estimated Lagrange multiplier}

\label{subsec:polarization}

In the special case $\muOm=\nuOm=0$, $\vec{v}=\vec{u}$, if we make
the identification $\lamstar=\vec{P}$ \eqref{eq:j} becomes identical
with the representation of $\vec{j}$ in terms of the electrostatic
dipole moment per unit volume or \emph{polarization vector} $\vec{P}$
{[}see e.g. §1-10 of \citet{Panofsky_Phillips_62}{]}. This representation
is as given in eq.~(12) of \citet{Calkin_63} and eq.~(1.2) of \citet{Webb_Anco_2017},
specialized to the MHD case of a \emph{quasineutral} moving medium,
where $\divv\vec{P}=0$ {[}consistently with \eqref{eq:divlamstar}{]}.

\citet{Calkin_63} goes on to derive an IMHD action principle in terms
of Clebsch potentials, but these are not globally defined in a 3-D
plasma with non-integrable magnetic fields. Our derivation shows the
Clebsch representation is not needed to apply this polarization representation
for $\vec{j}$ in an action principle if we apply the Lagrangian variational
approach of \citet{Newcomb_62}. {[}See also \citet{Webb_Anco_2019}
for a discussion of the equivalence of Lagrangian and Eulerian variational
approaches.{]}

\subsection{Explicit Lagrangian variation of fluid element position\label{subsec:Lagrangian-variation}}

This final Euler--Lagrange equation will in principle provide sufficient
equations to solve for the unknowns.

\subsubsection{Equations of motion}

From\eqref{eq:cderivs}\textbf{ $\partial_{\vec{v}}\cIOL=\Ident\cross\vec{B}$},
thus 
\begin{align}
\bm{\Pi} & =\rho\vec{u}+\Ident\cross\vec{B}\dotv\left(\lamIOL-\muIOL\,\cIOL\right)\;,\nonumber \\
 & =\rho\vec{u}+\vec{B}\cross\lamstar\;.\label{eq:PiDef-1}
\end{align}

The Euler--Lagrange equation obtained from setting $\delta\SQRx/\delta\vec{x}=0$
in \eqref{eq:xiEL} thus becomes 
\begin{align}
 & \partial_{t}\left(\rho\vec{u}+\vec{B}\cross\lamstar\right)+\divv\left[\vec{v}\left(\rho\vec{u}+\vec{B}\cross\lamstar\right)\right]+(\grad\vec{v})\dotv\left(\rho\vec{u}+\vec{B}\cross\lamstar\right)\nonumber \\
 & =\rho\grad\left(\frac{\partial\LdenQRx}{\partial\rho}\right)\,=\,\rho\grad\left(\vec{u}\dotv\vec{v}-h_{\Omega}\right)\;,\label{eq:xiEL-1}
\end{align}
by \eqref{eq:dLRxdrho}. Cancelling the $\rho(\grad\vec{v})\dotv\vec{u}$
occurring on both sides and rearranging, we have

\begin{align}
 & \rho\partial_{t}\vec{u}+\rho\vec{v}\dotv\grad\vec{u}-\rho\left(\grad\vec{u}\right)\dotv\vec{v}\nonumber \\
 & =-\rho\grad h_{\Omega}-\partial_{t}\left(\vec{B}\cross\lamstar\right)-\divv\left[\vec{v}\left(\vec{B}\cross\lamstar\right)\right]-(\grad\vec{v})\dotv\left(\vec{B}\cross\lamstar\right)\;,\label{eq:xiEL-2}
\end{align}
where we used \eqref{eq:Continuity}, $\partial_{t}\rho+\divv(\rho\vec{v})=0$,
to cancel all derivatives of $\rho$. Thus, dividing both sides by
$\rho$ we have the compact Bernoulli-like form

\begin{align}
\partial_{t}\vec{u}+\bm{\omega}\cross\vec{v} & =-\grad h_{\Omega}-\vec{a}_{\bm{\lambda}}\;,\label{eq:BernEqMot}
\end{align}
the residual acceleration term containing $\lamstar$ being 
\begin{align}
\vec{a}_{\bm{\lambda}}\defeq & \rho^{-1}\left[\partial_{t}\left(\vec{B}\cross\lamstar\right)+\divv\left[\vec{v}\left(\vec{B}\cross\lamstar\right)\right]+(\grad\vec{v})\dotv\left(\vec{B}\cross\lamstar\right)\right]\nonumber \\
= & \,\partial_{t}\vec{w}+\vec{v}\dotv\grad\vec{w}+(\grad\vec{v})\dotv\vec{w}\;,\label{eq:alambda}
\end{align}
where

\begin{align}
\vec{w}\,\defeq & \,\frac{\vec{B}\cross\lamstar}{\rho}\;,\label{eq:wdef}
\end{align}
again using $\partial_{t}\rho+\divv(\rho\vec{v})=0$.

In the special case $\muOm=\nuOm=0$, $\vec{v}=\vec{u}$ we can use
the identification in Subsection~\ref{subsec:polarization} of $\lamstar$
as the polarization field $\vec{P}$ to write $\vec{w}=\vec{B}\cross\vec{P}/{\rho}$.
We can then recognize \eqref{eq:BernEqMot} as the Eulerian equation
of motion, eq.~(23) of \citet{Calkin_63}, thus providing a physical
interpretation of our equations of motion in terms of a Lagrange multiplier
field.

\textbf{Check:} Calkin's (23) can be written as 
\begin{align*}
\partial_{t}\left(\vec{u}+\vec{w}\right)+\left[\curl\left(\vec{u}+\vec{w}\right)\right]\cross\vec{u} & =-\grad\left(h_{\Omega}+\vec{u}\dotv\vec{w}\right)\\
\text{i.e. }\partial_{t}\vec{u}+\bm{\omega}\cross\vec{u} & =-\grad h_{\Omega}-\grad\left(\vec{u}\dotv\vec{w}\right)\\
 & \qquad-\partial_{t}\vec{w}-\left(\curl\vec{w}\right)\cross\vec{u}\\
 & =-\grad h_{\Omega}-\vec{a}_{\vec{P}}\;,
\end{align*}
where
\begin{align*}
\vec{a}_{\vec{P}}\defeq & \partial_{t}\vec{w}+\left(\curl\vec{w}\right)\cross\vec{u}+\grad\left(\vec{u}\dotv\vec{w}\right)\\
= & \partial_{t}\vec{w}+\vec{u}\dotv\grad\vec{w}-\left(\grad\vec{w}\right)\dotv\vec{u}+\left(\grad\vec{u}\right)\dotv\vec{w}+\left(\grad\vec{w}\right)\dotv\vec{u}\\
= & \,\partial_{t}\vec{w}+\vec{u}\dotv\grad\vec{w}+(\grad\vec{u})\dotv\vec{w}\equiv\vec{a}_{\bm{\lambda}}\;\Box
\end{align*}

\subsubsection{Conservation form}

Now consider the conservation form \eqref{eq:MomConsvnForm} where,
from \eqref{eq:cderivs}, \eqref{eq:PiDef} and \eqref{eq:ConstraintLden}
\begin{align}
\bm{\Pi}+\epsF\frac{\partial\LdenQPr}{\partial\vec{E}}\cross\vec{B} & =\rho\vec{u}+\frac{\partial\LdenEXB}{\partial\vec{v}}+\epsF\frac{\partial\LdenEXB}{\partial\vec{E}}\cross\vec{B}\nonumber \\
 & =\rho\vec{u}+\frac{\partial\cIOL}{\partial\vec{v}}\dotv\lamstar+\epsF\left(\frac{\partial\cIOL}{\partial\vec{E}}\dotv\lamstar\right)\cross\vec{B}\\
 & =\rho\vec{u}+\left(1-\epsF\right)\vec{B}\cross\lamstar\;.\label{eq:PiEL}
\end{align}
Starting with the coefficient of $\vsf{I}$ in the tensor $\vsf{T}$,
\eqref{eq:Tdef}, and referring to \eqref{eq:RxPSL}, \eqref{eq:QRxPSL}
and \eqref{eq:LAminusdef} we find

\begin{align}
\LdenQPr-\rho\,\frac{\partial\LdenQPr}{\partial\rho} & =-\frac{p}{\gamma-1}-\frac{-\gamma\tau_{\Omega}\rho}{\gamma-1}-\frac{B^{2}}{2\muSI}+\nuOm\frac{\vec{u}\dotv\vec{B}}{\muSI}+\LdenEXB\nonumber \\
 & =p-\frac{B^{2}}{2\muSI}+\nuOm\frac{\vec{u}\dotv\vec{B}}{\muSI}+\lamIOL\dotv\cIOL-\frac{\muIOL\,\cIOL^{2}}{2}\;.
\end{align}

The penultimate term in $\vsf{T}$ is

\begin{align}
-\frac{\partial\LdenEXB}{\partial\vec{E}}\vec{E} & =-\lamstar\vec{E}
\end{align}
which consists of a symmmetric $\Ees\Ees$ term and a non-symmetric
$\Ees\vec{u}\cross\vec{B}$ term.

The preceding term of $\vsf T$ is, using \eqref{eq:dLCdB}and \eqref{eq:LemXIX},

\begin{align*}
\frac{\partial\LdenQPr}{\partial\vec{B}}\cross\Ident\cross\vec{B} & =\left(-\vec{B}+\nuOm\vec{u}-\muSI\vec{v}\cross\lamstar\right)\cross\Ident\cross\frac{\vec{B}}{\muSI}\\
 & =\frac{\vec{B}}{\muSI}\left(-\vec{B}+\nuOm\vec{u}-\muSI\vec{v}\cross\lamstar\right)\\
 & \quad-\frac{\Ident}{\muSI}\left(-B^{2}+\nuOm\vec{u}\dotv\vec{B}-\muSI\vec{v}\cross\lamstar\dotv\vec{B}\right)\;,
\end{align*}
and the remaining, first term is

\[
\vec{v}\bm{\Pi}=\left(\vec{u}-\frac{\nuOm\vec{B}}{\muSI\rho}\right)\left(\rho\vec{u}+\vec{B}\cross\lamstar\right)\;.
\]

Thus, combining all terms, \eqref{eq:MomConsvnForm} becomes

\begin{equation}
\partial_{t}\left(\rho\vec{u}\right)+\divv\left(\vsf T_{{\rm {\rm MHD}}}+\vsf T_{{\rm Res}}\right)=\left(\grad\lamIOL\right)\dotv\cIOL\;,\label{eq:MomConsvnForm-1}
\end{equation}
where $\vsf T_{{\rm {\rm MHD}}}$ is the momentum transport plus stress
tensor for both IMHD \emph{and} RxMHD,\citet{Dewar_Burby_Qu_Sato_Hole_2020},
\begin{equation}
\vsf T_{{\rm {\rm MHD}}}=\rho\vec{u}\vec{u}+\left(p+\frac{B^{2}}{2\muSI}\right)\Ident-\frac{\vec{B}\vec{B}}{\muSI}\;,
\end{equation}
the terms in $\nuOm$ that might have contributed to $\vsf T_{{\rm {\rm MHD}}}$
in the RxMHD case having cancelled.

The new term $\vsf T_{{\rm Res}}$ is the ``internal'' residual
stress contribution arising when action-extremizing solutions are
infeasible, i.e. when the IOL constraint is not satisfied exactly,

\begin{align}
\vsf T_{{\rm Res}} & \defeq\left(\lamIOL\dotv\cIOL-\frac{\muIOL\,\cIOL^{2}}{2}-\lamstar\dotv\vec{v}\cross\vec{B}\right)\Ident\nonumber \\
 & \quad-\vec{B}\vec{v}\cross\lamstar+\vec{v}\vec{B}\cross\lamstar-\lamstar\vec{E}\nonumber \\
 & =\left(\lamstar\dotv\cIOL-\lamstar\dotv\vec{u}\cross\vec{B}+\frac{\muIOL\,\cIOL^{2}}{2}\right)\Ident\nonumber \\
 & \quad+\vec{B}\lamstar\cross\vec{u}+\vec{u}\vec{B}\cross\lamstar+\lamstar\vec{u}\cross\vec{B}-\lamstar\vec{C}\;.\label{eq:TQRx}
\end{align}
(Interestingly, the $\nuOm$ cancellation also occurred in deriving
$\vsf T_{{\rm Res}}$ when $\vec{v}$ was replaced by $\vec{u}-\nuOm\vec{B}/\muSI\rho$.)

The ``external'' residual force on the RHS of \eqref{eq:MomConsvnForm-1}
obviously vanishes for feasible solutions. However it is not obvious
that $\vsf T_{{\rm Res}}$ vanishes when $\cIOL=0$ as it involves
the unknown converged Lagrange multiplier $\lamIOL|^{\infty}$ ($=\lamstar|^{\infty}$
as $\cIOL|^{\infty}=0$). However it is easy to verify that both the
diagonal and off-diagonal terms of $\vsf T_{{\rm Res}}$ not involving
$\cIOL$ explicitly do vanish if $\lamstar$ is proportional to $\vec{B}$
pointwise, implying at least in this case $\vsf T_{{\rm Res}}=0$
if and only if \textbf{$\cIOL=0$} (assuming $\muIOL\neq0$).

\subsubsection{Momentum and angular momentum conservation}

When a trial solution is IOL-infeasible, i.e. $\cIOL\neq0$, $\vsf T_{{\rm Res}}$
is not a symmetric tensor, indicating it imparts both an isotropic
pressure force and a torque on the plasma, presumably tending to change
$\vec{u}$ in such a way as to ``bend'' the flow toward conformity
with the Ideal Ohm's Law. There is a cyclic symmetry in $\vsf T_{{\rm Res}}$
among the three terms in $\lamstar,\vec{u},\vec{B}$ that indicate
that the magnetic field is coupled to $\Ees$ in a similar fashion
as $\vec{u}$, and indeed we see from \eqref{eq:j} that that there
is a ``dynamo'' term depending on $\Ees$ in $\vec{j}$ that modifies
$\vec{B}$, by Ampère's Law.

\section{Linearized dynamics in the WKB approximation \label{sec:WKBRxMHD}}

As indicated in the Introduction, the present paper is a step toward
a multi-region RxMHD dynamics code in which the primary role of the
relaxed fluid dynamics within an annular toroidal domain $\Omega$
is twofold a) to regularize IMHD by relaxing the topological constraint
forbidding magnetic reconnection, so magnetic islands can form at
resonances rather than singularities, and b) to transmit pressure
disturbances across the thin layer of plasma between the two disjoint
interfaces forming the boundary $\partial\Omega$, thereby coupling
the interfaces and endowing them with the plasma's inertia. This section
derives, in the WKB approximation, dispersion relations for the waves
that transmit these disturbances.

\subsection{Linearization \label{subsec:LinRxMHD} }

Thus, as a simple first step toward understanding the dynamical implications
of the RxMHD equations we linearize around a steady, ($\partial_{t}\mapsto0$),
IOL-compliant $\left(C^{(0)}=0,\,\lamstar^{(0)}=\lamIOL^{(0)}\right)$
solution of the Euler--Lagrange equations in a domain $\Omega$ with
either fixed boundaries or with only low-amplitude, short-wavelength
perturbations on $\partial\Omega$. Thus, insert in these equations
the ansatz $\vec{u}=\vec{u}^{(0)}+\alpha\vec{u}^{(1)}+O(\alpha^{2})$,
where $\alpha$ is the amplitude expansion parameter, and similarly
for other perturbations except we use their potential representations
for $\vec{B}^{(1)}$ and \textbf{$\vec{E}^{(1)}$} as this is important
for enforcing \ref{eq:MaxFad}. The entropy, helicity and cross-helicity
integrals are conserved at $O(\alpha)$, with therefore no perturbation
in the Lagrange multipliers. Thus here we take $\tau_{\Omega}$, $\mu_{\Omega}$,
and $\nuOm$ as time-independent constants. Also, from here on we
take the superscript (0) to be implicit, e.g. $\rho$ means $\rho^{(0)}$,
$\vec{u}$ means $\vec{u}^{(0)}$, $\lamIOL$ means $\lamIOL^{(0)}$
etc. While we assume the background equilibrium obeys the IOL, we
do not assume the augmented-Lagrangian iteration for our perturbations
is fully converged, so $\cIOL^{(1)}\neq0$ and our two successive
Euler-Lagrange iterates are not equal, $\lamstar^{(1)}\neq\lamIOL^{(1)}$.

\subsubsection{Linearization of Lagrange multiplier determination \label{subsec:Linearization-of-Lagrange}}

Focusing first on the novel part of the calculation we list the linearizations
of immediate relevance to the Augmented Lagrangian determination of
the updated Lagrange multiplier field $\lamstar$.

From \eqref{eq:Brep} and \eqref{eq:Erep}, $\vec{B}^{(1)}=\curl\vec{A}^{(1)}$
and $\vec{E}^{(1)}=-\grad\Phi^{(1)}-\epsF\partial_{t}\vec{A}^{(1)}$,
so \eqref{eq:cdef} becomes 
\begin{equation}
\cIOL^{(1)}=-\grad\Phi^{(1)}-\epsF\partial_{t}\vec{A}^{(1)}+\vec{u}\cross\left(\curl\vec{A}^{(1)}\right)+\vec{u}^{(1)}\cross\vec{B}\label{eq:c-lin}
\end{equation}
with $\Phi^{(1)}$ to be determined from \eqref{eq:PoissonEq}, which
used the Euler--Lagrange equation from the $\Phi$ variation in its
derivation. This becomes

\begin{align}
\nabla^{2}\Phi^{(1)} & =\divv\left[\vec{u}^{(1)}\cross\vec{B}+\vec{u}\cross\left(\curl\vec{A}^{(1)}\right)\right]\;.\label{eq:PoissonEq-lin}
\end{align}
When $\cIOL^{(1)}$ is found, the updated Lagrange multiplier is determined
from the linearization of \eqref{eq:lamstardef}, $\lamstar^{(1)}=\lamIOL^{(1)}-\muIOL\cIOL^{(1)}$.

While one occurrence of $\vec{A}^{(1)}$ in \eqref{eq:PoissonEq-lin}
has been eliminated by assuming Coulomb gauge, $\divv\vec{A}^{(1)}=0$,
it still arises in the $\curl\vec{A}^{(1)}$ term arising from $\vec{B}^{(1)}$.
Thus we also need the linearization of the $\delta\vec{A}$ Euler--Lagrange
equation to give us $\vec{A}^{(1)}$. We use the modified Beltrami
form \eqref{eq:ModBel} 
\begin{align}
\curl\left(\curl\vec{A}^{(1)}\right) & =\mu_{\Omega}\curl\vec{A}^{(1)}+\nuOm\curl\vec{u}^{(1)}-\muSI\curl\left(\vec{v}^{(1)}\cross\lamIOL\right)\nonumber \\
 & \quad+\muSI\left[\epsF\partial_{t}\lamstar^{(1)}-\curl\left(\vec{v}\cross\lamstar^{(1)}\right)\right]\;,\label{eq:ModBel-lin}
\end{align}
where 
\begin{equation}
\vec{v}^{(1)}=\vec{u}^{(1)}-\frac{\nuOm}{\muSI}\left(\frac{\curl\vec{A}^{(1)}}{\rho}-\frac{\rho^{(1)}}{\rho}\frac{\vec{B}}{\rho}\right)\;,\label{eq:deltauEL-lin}
\end{equation}
with $\rho^{(1)}$ to be determined from 
\begin{equation}
\partial_{t}\rho^{(1)}+\divv(\rho\vec{u}^{(1)}+\rho^{(1)}\vec{u})=0\;,\label{eq:ucont-lin}
\end{equation}
and $\vec{u}^{(1)}$ to be treated as the one unknown in terms of
which all other physical perturbations are to be expressed.

\subsection{Wave perturbations in WKB approximation}

\subsubsection{Eikonal ansatz and natural basis vectors\label{subsec:Eikonal-ansatz}}

For short wavelength, high frequency velocity perturbations we use
the eikonal ansatz 
\begin{equation}
\vec{u}^{(1)}=\widetilde{\vec{u}}(\vec{x},t)\exp\left(\frac{\ij\varphi(\vec{x},t)}{\varepsilon}\right)\;,\label{eq:eikonvRxHD}
\end{equation}
with similar notations for linear perturbations of other quantities,
$\varepsilon$ being the WKB (local plane-wave) expansion parameter.
The instantaneous local values of wave vector and frequency as seen
in the LAB frame are then defined as $\vec{k}\defeq\grad\varphi$
and $\omega(\vec{x},t)\defeq-\partial_{t}\varphi$.

In the following development we shall also encounter two ``shifted''
frequencies: 
\begin{equation}
\text{(1)}\qquad\omega_{\vec{k}}^{\prime\vec{u}}\defeq\omega-\vec{k}\dotv\vec{u}\label{eq:omegakprimedef}
\end{equation}
the Doppler-shifted frequency of the wave as seen in the local rest
frame of a fluid element, velocity $\vec{u}(\vec{x},t)$, and 
\begin{equation}
\text{(2)}\qquad\omega_{\vec{k}}^{\prime\vec{v}}\defeq\epsF\omega-\vec{k}\dotv\vec{v}\label{eq:omegavkprimedef}
\end{equation}
the same as frequency (1) except with $\vec{u}$ replaced by the relative
velocity $\vec{v}=\vec{u}-\uRx\equiv\vec{u}-\nuOm\vec{B/}\muSI\rho$.

Taking $\varphi$ and equilibrium quantities to vary on $O(1)$ spatial
and temporal scales, $\omega$, $\vec{k}$, $\partial_{t}\vec{u}$,
$\grad\vec{u}$, $\mu_{\Omega}$, $\nuOm$ etc. are $O(1)$, but $\partial_{t}\vec{u}^{(1)}$,
$\grad\vec{u}^{(1)}$ etc. are large, $O(\alpha\varepsilon^{-1})$,
relative to $\vec{u}^{(1)}=O(\alpha)\varepsilon^{0}$, and similarly
for spatio-temporal derivatives of $\rho^{(1)}$ and $\vec{B}^{(1)}$.

In order for $\vec{B}^{(1)}$ to be the same order as $\vec{u}^{(1)}$,
the potentials $\Phi^{(1)}$ and$\vec{A}^{(1)}$ must be $O(\alpha\varepsilon)$,
so we write

\begin{equation}
\Phi^{(1)}=\frac{\varepsilon}{\ij}\widetilde{\Phi}(\vec{x},t)\exp\left(\frac{\ij\varphi(\vec{x},t)}{\varepsilon}\right)\;\text{and\;}\vec{A}^{(1)}=\frac{\varepsilon}{\ij}\widetilde{\vec{A}}(\vec{x},t)\exp\left(\frac{\ij\varphi(\vec{x},t)}{\varepsilon}\right)\;,\label{eq:eikonEM}
\end{equation}
\eqref{eq:Brep} and \eqref{eq:Erep} giving then $\widetilde{\vec{B}}=\vec{k}\cross\widetilde{\vec{A}}$
and $\widetilde{\vec{E}}=-\vec{k}\,\widetilde{\Phi}+\epsF\omega\,\widetilde{\vec{A}}$.

As in \citet{Dewar_Burby_Qu_Sato_Hole_2020} our strategy is to express
all perturturbations in terms of $\widetilde{\vec{u}}$, in order
to find a $3\times3$ matrix eigenvalue equation whose roots give
the dispersion relations for the three propagating wave branches.
We shall also include a forcing term of the form similar to \eqref{eq:eikonvRxHD}
in the equation of motion for $\vec{u}^{(1)}$ so that this $3\times3$
matrix appears also as a response function, with the dispersion relations
giving the location of its poles.

We shall find it useful to expand vectors and dyadics in the orthonormal
MHD-wave basis 
\begin{equation}
\esub{1}\defeq\frac{\vec{k}_{\perp}}{k_{\perp}}\;,\;\esub{2}\defeq\frac{\vec{B}}{B}\equiv\esub{\vec{B}}\;\;\text{and}\;\;\esub{3}\defeq\equiv\esub{1}\cross\esub{2}\equiv\frac{\vec{k}\cross\vec{B}}{\left|\vec{k}\cross\vec{B}\right|}\;,\label{eq:waveBasis}
\end{equation}
where $\vec{k}_{\perp}\defeq\Pperp\dotv\vec{k}$, so $\vec{\ensuremath{k}}=k_{\perp}\esub{1}+k_{\parallel}\esub{2}$
(where $k_{\perp}\defeq\left|\vec{k}_{\perp}\right|$, $k_{\parallel}\defeq\vec{k}\dotv\esub{\vec{B}}$),
$\uRx=u^{{\rm Rx}}\esub{2}$ (where $u^{{\rm Rx}}\defeq\left|\uRx\right|$),
$\vec{k}\cross\vec{B}_{{\rm A}}=k_{\perp}c_{{\rm A}}\esub{3}$, $\Pperp=\esub{1}\esub{1}+\esub{3}\esub{3}$
and $\Ident=\esub{1}\esub{1}+\esub{2}\esub{2}+\esub{3}\esub{3}$.
(There is of course a problem if $\vec{k}_{\perp}=0$, but we are
interested in low-frequency MHD waves around $k_{\parallel}=0$ where
$\left|k_{\perp}\right|$is maximal.)

\subsubsection{IOL Constraint in WKB approximation \label{subsec:WKB_IOLRxMHD}}

We now use \eqref{eq:eikonvRxHD} and \eqref{eq:eikonEM} in the linearizations
in Subsection~\ref{subsec:Linearization-of-Lagrange}, working to
leading order in $\varepsilon$(for instance $\mu_{\Omega}\vec{B}^{(1)}$
will be dropped as higher order in $\varepsilon$ than other terms
in \eqref{eq:ModBel-lin}). Then \eqref{eq:c-lin} becomes

\begin{align}
\widetilde{\cIOL} & =-\vec{k}\,\widetilde{\Phi}+\epsF\,\omega\widetilde{\vec{A}}+\vec{u}\cross\left(\vec{k}\cross\widetilde{\vec{A}}\right)+\widetilde{\vec{u}}\cross\vec{B}\nonumber \\
 & =-\vec{k}\left(\widetilde{\Phi}-\vec{u}\dotv\widetilde{\vec{A}}\right)+\omega_{\vec{k}}^{\prime\vec{u}}\widetilde{\vec{A}}+\widetilde{\vec{u}}\cross\vec{B}\;,\label{eq:c-tilde}
\end{align}
Also \eqref{eq:PoissonEq-lin} becomes

\begin{align}
\vec{k}^{2}\,\widetilde{\Phi} & =\vec{k}\dotv\widetilde{\vec{u}}\cross\vec{B}+\vec{k}\dotv\vec{u}\cross\left(\vec{k}\cross\widetilde{\vec{A}}\right)\;,\nonumber \\
 & =\vec{k}^{2}\vec{u}\dotv\widetilde{\vec{A}}+\vec{k}\dotv\widetilde{\vec{u}}\cross\vec{B}\;,\label{eq:PoissonEq-tilde}
\end{align}
assuming Coulomb gauge, $\vec{k}\dotv\widetilde{\vec{A}}=0$. Inserting
\eqref{eq:PoissonEq-tilde} in \eqref{eq:c-tilde} gives 
\begin{equation}
\widetilde{\cIOL}=\left(\Ident-\frac{\vec{k}\vec{k}}{\vec{k}^{2}}\right)\dotv\widetilde{\vec{u}}\cross\vec{B}+\omega_{\vec{k}}^{\prime\vec{u}}\widetilde{\vec{A}}\;\label{eq:c-tilde-1}
\end{equation}

Next, \eqref{eq:ModBel-lin} becomes

\begin{align}
\vec{k}^{2}\widetilde{\vec{A}} & =-\nuOm\vec{k}\cross\widetilde{\vec{u}}+\muSI\vec{k}\cross\left(\widetilde{\vec{v}}\cross\lamIOL\right)\nonumber \\
 & \quad+\muSI\left[\epsF\omega\,\widetilde{\lamstar}+\vec{k}\cross\left(\vec{v}\cross\widetilde{\lamstar}\right)\right]\;,\nonumber \\
 & =-\nuOm\vec{k}\cross\widetilde{\vec{u}}-\muSI\vec{k}\dotv\widetilde{\vec{v}}\,\lamIOL+\muSI\omega_{\vec{k}}^{\prime\vec{v}}\,\widetilde{\lamstar}\;,\label{eq:A-tilde}
\end{align}
where we used $\vec{v}=\vec{u}-\uRx\equiv\vec{u}-\nuOm\vec{B/}\muSI\rho$.

Finally, \eqref{eq:ucont-lin} and \eqref{eq:deltauEL-lin} become

\begin{align}
\frac{\widetilde{\rho}}{\rho} & =\frac{\vec{k}\dotv\widetilde{\vec{u}}}{\omega_{\vec{k}}^{\prime\vec{u}}}\;,\label{eq:rho-tilde}\\
\widetilde{\vec{v}} & =\widetilde{\vec{u}}+\frac{\nuOm}{\muSI\rho}\left(\frac{\vec{k}\dotv\widetilde{\vec{u}}}{\omega_{\vec{k}}^{\prime\vec{u}}}\vec{B}-\vec{k}\cross\widetilde{\vec{A}}\right)\label{eq:v-tilde}
\end{align}
hence 
\begin{align}
\vec{k}\dotv\widetilde{\vec{v}} & =\left(1+\frac{\vec{k}\dotv\uRx}{\omega_{\vec{k}}^{\prime\vec{u}}}\right)\vec{k}\dotv\widetilde{\vec{u}}\;.\nonumber \\
 & =\left(\frac{\omega_{\vec{k}}^{\prime\vec{v}}}{\omega_{\vec{k}}^{\prime\vec{u}}}\right)\vec{k}\dotv\widetilde{\vec{u}}\;.\label{eq:kdotv-tilde}
\end{align}

Substituting\eqref{eq:kdotv-tilde} in \eqref{eq:A-tilde} gives

\begin{align}
\vec{k}^{2}\widetilde{\vec{A}} & =-\nuOm\vec{k}\cross\widetilde{\vec{u}}-\muSI\left(\frac{\omega_{\vec{k}}^{\prime\vec{v}}}{\omega_{\vec{k}}^{\prime\vec{u}}}\right)\vec{k}\dotv\widetilde{\vec{u}}\,\lamIOL+\muSI\omega_{\vec{k}}^{\prime\vec{v}}\,\widetilde{\lamstar}\;,\label{eq:A-tilde-1}
\end{align}
which in \eqref{eq:c-tilde-1} then gives

\begin{align}
\widetilde{\cIOL} & =\left(\Ident-\frac{\vec{k}\vec{k}}{\vec{k}^{2}}\right)\dotv\widetilde{\vec{u}}\cross\vec{B}-\nuOm\omega_{\vec{k}}^{\prime\vec{u}}\frac{\vec{k}\cross\widetilde{\vec{u}}}{\vec{k}^{2}}+\muSI\frac{\omega_{\vec{k}}^{\prime\vec{v}}}{\vec{k}^{2}}\left(\omega_{\vec{k}}^{\prime\vec{u}}\,\widetilde{\lamstar}-\vec{k}\dotv\widetilde{\vec{u}}\,\lamIOL\right)\;.\label{eq:c-tilde-2}
\end{align}

Treating $\widetilde{\cIOL}$ for the moment as a known and solving
for $\widetilde{\lamstar}$ we have

\begin{align}
\widetilde{\lamstar} & =\frac{\vec{k}\dotv\widetilde{\vec{u}}}{\omega_{\vec{k}}^{\prime\vec{u}}}\,\lamIOL+\frac{\nuOm}{\muSI}\;\frac{\vec{k}\cross\widetilde{\vec{u}}}{\omega_{\vec{k}}^{\prime\vec{v}}}+\frac{\left(\vec{k}^{2}\Ident-\vec{k}\vec{k}\right)\cross\vec{B}\dotv\widetilde{\vec{u}}+\vec{k}^{2}\widetilde{\cIOL}}{\muSI\,\omega_{\vec{k}}^{\prime\vec{u}}\,\omega_{\vec{k}}^{\prime\vec{v}}}\;.\label{eq:lamstar-tilde}
\end{align}

Normally we do not need to know the residual IOL error term $\widetilde{\cIOL}$
exactly, but to convince ourselves it can be made arbitrarily small
by iteration, replace $\widetilde{\lamstar}$ with its explicit form
from the linearization of \eqref{eq:lamstardef}, $\widetilde{\lamIOL}-\muIOL\widetilde{\cIOL}$,
and collect both $\widetilde{\cIOL}$ terms on the left:

\begin{align}
\left(1+\muSI\omega_{\vec{k}}^{\prime\vec{u}}\omega_{\vec{k}}^{\prime\vec{v}}\,\frac{\muIOL}{\vec{k}^{2}}\right)\widetilde{\cIOL} & =-\left(\Ident-\frac{\vec{k}\vec{k}}{\vec{k}^{2}}\right)\cross\vec{B}\dotv\widetilde{\vec{u}}-\nuOm\omega_{\vec{k}}^{\prime\vec{u}}\frac{\vec{k}\cross\widetilde{\vec{u}}}{\vec{k}^{2}}\nonumber \\
 & \quad+\muSI\frac{\omega_{\vec{k}}^{\prime\vec{v}}}{\vec{k}^{2}}\left(\omega_{\vec{k}}^{\prime\vec{u}}\,\widetilde{\lamIOL}-\vec{k}\dotv\widetilde{\vec{u}}\,\lamIOL\right)\;,\label{eq:c-tilde-3}
\end{align}
which confirms the implication in Subsection~\eqref{eq:QEXB} that
we have enough equations to determine $\widetilde{\cIOL}$, and hence
$\widetilde{\Phi}$, $\widetilde{\vec{A}}$ and $\widetilde{\lamstar}$,
in terms of $\widetilde{\vec{u}}$.)

Dividing both sides of \eqref{eq:c-tilde-3} by the large penalty
multiplier $\muIOL$ we see that $\widetilde{\cIOL}$ is smaller than
the other terms and the previous iterate of $\widetilde{\cIOL}$ by
an $O\left(1/\muIOL\right)$ factor. Thus the sequence $\left\{ \ldots,\widetilde{\cIOL}^{n},\widetilde{\cIOL}^{n+1},\widetilde{\cIOL}^{n+2},\ldots\right\} $
will converge exponentially toward $0$, or super-exponentially if
$\muIOL$ is increased appropriately at each step. Also $\widetilde{\lamstar}$
will converge to $\widetilde{\lamIOL}|^{\infty}$.

However, this linearized calculation is sufficiently simple that we
do not actually need to carry out the iteration as we can find $\widetilde{\lamIOL}|^{\infty}$analytically
from \ref{eq:c-tilde-3} by setting its LHS to zero and solving for
$\widetilde{\lamIOL}=\widetilde{\lamIOL}|^{\infty}$. Or, if we want
to investigate the hypothesis that terminating the iteration at finite
$n$, so that $\widetilde{\cIOL}\neq0,$ will regularize MHD we can
\emph{prescribe} $\widetilde{\cIOL}$ and use \eqref{eq:lamstar-tilde}
to give $\widetilde{\lamstar}$. To provide a continuous sequence
of dynamical fluid models running between the unconstrained RxMHD
perturbation dynamics of \citet{Dewar_Burby_Qu_Sato_Hole_2020} to
the converged, $\widetilde{\cIOL}=0$, present model we choose 
\begin{equation}
\widetilde{\cIOL}=\epsRx\widetilde{\cIOL}_{0}\;,\label{eq:cSequence}
\end{equation}
with the ``relaxedness'' parameter $\epsRx$ running from 0 (IMHD,
IOL-compliant) to 1 (RxMHD, may be IOL-infeasible). We shall later
also have use of the complementary ``ideality'' parameter $\epsI\defeq1-\epsRx$.

Here $\widetilde{\cIOL}_{0}$ is the IOL error for unconstrained,
$\epsI=0,\;\epsRx=1$, RxMHD perturbations, which we can find by setting
$\widetilde{\lamIOL}=0$, $\muIOL=0$, and thus $\widetilde{\lamstar}=0$,
in \eqref{eq:c-tilde-2} to give 
\begin{align}
\widetilde{\cIOL_{0}} & =\left(\Ident-\frac{\vec{k}\vec{k}}{\vec{k}^{2}}\right)\dotv\widetilde{\vec{u}}\cross\vec{B}-\nuOm\omega_{\vec{k}}^{\prime\vec{u}}\frac{\vec{k}\cross\widetilde{\vec{u}}}{\vec{k}^{2}}-\muSI\lamIOL\omega_{\vec{k}}^{\prime\vec{v}}\frac{\vec{k}\dotv\widetilde{\vec{u}}}{\vec{k}^{2}}\;,\label{eq:c-tilde-4}
\end{align}
which is a linear tensor function of the form $\widetilde{\cIOL_{0}}=B\,\vsf{C}_{0}(\vec{k},\vec{B})\dotv\widetilde{\vec{u}}$,
where the factor $B$ is taken out to make $\vsf{C}_{0}$ dimensionless.
By inspection of \eqref{eq:c-tilde-4}, 
\begin{equation}
\vsf{C}_{0}\defeq-\frac{1}{\vec{k}^{2}B}\left[\muSI\omega_{\vec{k}}^{\prime\vec{v}}\lamIOL\vec{k}+\nuOm\omega_{\vec{k}}^{\prime\vec{u}}\vec{k}\cross\Ident+\left(\vec{k}^{2}\Ident-\vec{k}\vec{k}\right)\cross\vec{B}\right]\;.\label{eq:tensorrepC0}
\end{equation}

The linear tensor form of $\widetilde{\cIOL_{0}}$ implies $\widetilde{\cIOL}$,
$\widetilde{\lamIOL}$ and $\widetilde{\lamstar}$ are of similar
form, 
\begin{equation}
\widetilde{\cIOL}=B\,\vsf{C}(\vec{k},\vec{B})\dotv\widetilde{\vec{u}},\quad\widetilde{\lamIOL}=\vsf{\Lambda}(\vec{k},\vec{B})\dotv\widetilde{\vec{u}},\quad\widetilde{\lamstar}=\vsf{\Lambda}_{*}(\vec{k},\vec{B})\dotv\widetilde{\vec{u}}\;,\label{eq:tensorrep}
\end{equation}
where from $\vsf{\Lambda}=\vsf{\Lambda}_{*}+\muIOL B\,\vsf{C}$, and
from \eqref{eq:lamstar-tilde} and \eqref{eq:tensorrep}, 
\begin{align}
\vsf{\Lambda}_{*} & =\frac{\lamIOL\,\vec{k}}{\omega_{\vec{k}}^{\prime\vec{u}}}\,+\frac{\nuOm}{\muSI}\;\frac{\vec{k}\cross\Ident}{\omega_{\vec{k}}^{\prime\vec{v}}}+\frac{\left(\vec{k}^{2}\Ident-\vec{k}\vec{k}\right)\cross\vec{B}+\vec{k}^{2}B\vsf{C}}{\muSI\,\omega_{\vec{k}}^{\prime\vec{u}}\,\omega_{\vec{k}}^{\prime\vec{v}}}\nonumber \\
 & =\epsI\left[\frac{\lamIOL\,\vec{k}}{\omega_{\vec{k}}^{\prime\vec{u}}}\,+\frac{\nuOm}{\muSI}\;\frac{\vec{k}\cross\Ident}{\omega_{\vec{k}}^{\prime\vec{v}}}+\frac{\left(\vec{k}^{2}\Ident-\vec{k}\vec{k}\right)\cross\vec{B}}{\muSI\,\omega_{\vec{k}}^{\prime\vec{u}}\,\omega_{\vec{k}}^{\prime\vec{v}}}\right]\;\text{,}\label{eq:lamstartensor}
\end{align}
using $\vsf{C}=\epsRx\vsf{C}_{0}$, \ref{eq:cSequence}.

\subsubsection{Short-wavelength dynamical RxMHD equations}

We now consider the linearized equation of motion with the forcing
term mentioned in Subsection~\ref{subsec:Eikonal-ansatz}, a specific
force (i.e. force/mass density) we denote as $\vec{f}^{(1)}$. Thus
the linearized \eqref{eq:BernEqMot} with forcing term becomes

\begin{align}
\partial_{t}\vec{u}^{(1)}+\bm{\omega}^{(1)}\cross\vec{v} & +\bm{\omega}\cross\vec{v}^{(1)}=-\vec{a}_{\bm{\lambda}}^{(1)}-\grad h_{\Omega}^{(1)}+\vec{f}^{(1)}\;,\label{eq:BernEqMot-1}
\end{align}
where, from \eqref{eq:Head}, 
\begin{equation}
\begin{split}h_{\Omega}^{(1)} & =\vec{u}\dotv\vec{u}^{(1)}+\text{\ensuremath{\tau_{\Omega}}}\frac{\rho^{(1)}}{\rho}\;,\end{split}
\label{eq:Head-lin}
\end{equation}
from \eqref{eq:alambda},

\begin{align}
\vec{a}_{\bm{\lambda}}^{(1)}= & \partial_{t}\vec{w}^{(1)}+\vec{v}\dotv\grad\vec{w}^{(1)}+(\grad\vec{v})\dotv\vec{w}^{(1)}\nonumber \\
 & \,\qquad+\vec{v}^{(1)}\dotv\grad\vec{w}+\left(\grad\vec{v}{}^{(1)}\right)\dotv\vec{w},\label{eq:alambda-lin}
\end{align}
and, from \eqref{eq:wdef}, 
\begin{align}
\vec{w}^{(1)}=\, & \frac{\vec{B}^{(1)}\cross\lamIOL}{\rho}+\frac{\vec{B}\cross\lamstar^{(1)}}{\rho}-\frac{\rho^{(1)}\vec{B}\cross\lamIOL}{\rho^{2}}\;.\label{eq:w-lin}
\end{align}

Using the WKB representations in \eqref{eq:eikonEM} and in \eqref{eq:eikonvRxHD},
and analogous representations for $\rho^{(1)}$, $\vec{B}^{(1)}$,
$\vec{w}^{(1)}$, $\vec{a}_{\bm{\lambda}}^{(1)}$ and $\vec{f}^{(1)}$
in the linearizations above, and working to leading order in $\varepsilon$
as before we have

\begin{align}
-\omega\widetilde{\vec{u}}+\left(\vec{k}\cross\widetilde{\vec{u}}\right)\cross\vec{v} & =-\vec{k}\left(\vec{u}+\text{\ensuremath{\tau_{\Omega}}}\frac{\vec{k}}{\omega_{\vec{k}}^{\prime\vec{u}}}\right)\dotv\widetilde{\vec{u}}-\widetilde{\vec{a}}_{\bm{\lambda}}+\widetilde{\vec{f}}\;,\label{eq:BernEqMot-tilde}
\end{align}
which we shall show can be written as 
\begin{equation}
\vsf{D}\left(\omega,\vec{k}\right)\dotv\widetilde{\vec{u}}=-\widetilde{\vec{f}}\label{eq:u-tildeDrive}
\end{equation}
where, noting canceling of $\vec{k}\vec{u}\dotv\widetilde{\vec{u}}$
terms,

\begin{align}
\vsf{D}\dotv\widetilde{\vec{u}} & =\omega\widetilde{\vec{u}}-\vec{k}\dotv\vec{u}\,\widetilde{\vec{u}}-\uRx\cross\left(\vec{k}\cross\widetilde{\vec{u}}\right)-\widetilde{\vec{a}}_{\bm{\lambda}}-\text{\ensuremath{\tau_{\Omega}}}\frac{\vec{k}\vec{k}}{\omega_{\vec{k}}^{\prime\vec{u}}}\dotv\widetilde{\vec{u}}\label{eq:Ddef}
\end{align}

From \eqref{eq:alambda-lin},

\begin{align}
\widetilde{\vec{a}}_{\bm{\lambda}}= & -\omega_{\vec{k}}^{\prime\vec{v}}\,\widetilde{\vec{w}}+\vec{k}\widetilde{\vec{v}}\dotv\vec{w}\;.\label{eq:alambda-tilde}
\end{align}
and, from \eqref{eq:w-lin}, \eqref{eq:rho-tilde} and \eqref{eq:A-tilde-1},

\begin{align}
\rho\widetilde{\vec{w}}=\, & \left(\vec{k}\cross\widetilde{\vec{A}}\right)\cross\lamIOL-\frac{\widetilde{\rho}}{\rho}\vec{B}\cross\lamIOL+\vec{B}\cross\widetilde{\lamstar}\nonumber \\
=\, & \left[\frac{\vec{k}}{\vec{k}^{2}}\cross\left(-\nuOm\vec{k}\cross\widetilde{\vec{u}}-\muSI\left(\frac{\omega_{\vec{k}}^{\prime\vec{v}}}{\omega_{\vec{k}}^{\prime\vec{u}}}\right)\vec{k}\dotv\widetilde{\vec{u}}\,\lamIOL\right)-\frac{\vec{k}\dotv\widetilde{\vec{u}}}{\omega_{\vec{k}}^{\prime\vec{u}}}\vec{B}\right]\cross\lamIOL\nonumber \\
+ & \left[\frac{\muSI\omega_{\vec{k}}^{\prime\vec{v}}}{\vec{k}^{2}}\left(\lamIOL\dotv\vec{k}\,\Ident-\vec{k}\lamIOL\right)+\vec{B}\cross\Ident\right]\dotv\vsf{\Lambda}_{*}\dotv\widetilde{\vec{u}}\;,\label{eq:w-tilde}
\end{align}
where $\vsf{\Lambda}_{*}$ was given in \eqref{eq:lamstartensor}.

\subsubsection{WKB RxMHD response matrix \label{subsec:WKB_Response}}

To simplify the calculation of the response we expand around a relaxed
equilibrium with $\lamIOL=0$, such as the axisymmetric tokamak equilibrium
in \citet{Dewar_Burby_Qu_Sato_Hole_2020}, which has a steady flow
field $\vec{u}$ that is the vector sum of an arbitrary rigid toroidal
rotation carried by $\vec{v}$ and an axisymmetric magnetic-field-aligned
flow $\uRx$ proportional to $\nuOm$, which equilibrium was shown
to satisfy the IOL without needing a Lagrange multiplier.

In such a case $\vec{w}=0$ and \eqref{eq:w-tilde} becomes $\widetilde{\vec{w}}=\vec{B}\cross\vsf{\Lambda}_{*}\dotv\widetilde{\vec{u}}/\rho$.
Then \eqref{eq:alambda-tilde} becomes

\begin{align}
\widetilde{\vec{a}}_{\bm{\lambda}}= & -\epsI\frac{\vec{B}}{\rho}\cross\left[\frac{\nuOm}{\muSI}\;\vec{k}\cross\Ident+\frac{\left(k^{2}\Ident-\vec{k}\vec{k}\right)\cross\vec{B}}{\muSI\,\omega_{\vec{k}}^{\prime\vec{u}}\,}\right]\dotv\widetilde{\vec{u}}\nonumber \\
= & -\epsI\uRx\cross\left(\vec{k}\cross\widetilde{\vec{u}}\right)-\frac{\epsI\vec{B}\cross\left(k^{2}\Ident-\vec{k}\vec{k}\right)\cross\vec{B}}{\muSI\rho\,\omega_{\vec{k}}^{\prime\vec{u}}}\dotv\widetilde{\vec{u}}\;.\label{eq:alambda-tilde-1}
\end{align}

In \eqref{eq:u-tildeDrive} the $\uRx$ terms not involving $\epsRx$
cancel, giving

\begin{align}
\vsf{D} & =\omega_{\vec{k}}^{\prime\vec{u}}\Ident+\frac{\epsI\vec{B}\cross\left(k^{2}\Ident-\vec{k}\vec{k}\right)\cross\vec{B}}{\muSI\rho\,\omega_{\vec{k}}^{\prime\vec{u}}}-\epsRx\left(\vec{k}\uRx-\vec{k}\dotv\uRx\Ident\right)-\text{\ensuremath{\tau_{\Omega}}}\frac{\vec{k}\vec{k}}{\omega_{\vec{k}}^{\prime\vec{u}}}\nonumber \\
 & =\left(\omega_{\vec{k}}^{\prime\vec{u}}+\epsRx\vec{k}\dotv\uRx\right)\Ident-\epsRx\vec{k}\,\uRx+\epsI\frac{\vec{c}_{{\rm A}}\cross\left(k^{2}\Ident-\vec{k}\vec{k}\right)\cross\vec{c}_{{\rm A}}}{\omega_{\vec{k}}^{\prime\vec{u}}}-c_{{\rm s}}^{2}\frac{\vec{k}\vec{k}}{\omega_{\vec{k}}^{\prime\vec{u}}}\nonumber \\
 & =\left(\omega_{\vec{k}}^{\prime\vec{u}}+\epsRx\vec{k}\dotv\uRx\right)\Ident-\epsRx\vec{k}\,\uRx+\epsI\frac{\vec{k}_{\perp}\cross\vec{c}_{{\rm A}}\,\vec{k}_{\perp}\cross\vec{c}_{{\rm A}}-k^{2}c_{{\rm A}}^{2}\Pperp}{\omega_{\vec{k}}^{\prime\vec{u}}}\nonumber \\
 & \qquad\qquad\qquad\qquad\qquad\qquad\qquad\qquad-c_{{\rm s}}^{2}\frac{\vec{k}\vec{k}}{\omega_{\vec{k}}^{\prime\vec{u}}}\;,\label{eq:Ddef-1}
\end{align}
where $\uRx=\nuOm\vec{B}/\muSI\rho$ is defined in \eqref{eq:ufullRx},
$\vec{c}_{{\rm A}}\defeq\vec{B}/\left(\muSI\rho\right)^{1/2}$ is
the Alfvén velocity, $c_{{\rm s}}=\text{\ensuremath{\tau_{\Omega}^{1/2}}}$
is the isothermal sound speed, and we have used \eqref{eq:LemXIX}
to write $\vec{c}_{{\rm A}}\cross\Ident\cross\vec{c}_{{\rm A}}=\vec{c}_{{\rm A}}\vec{c}_{{\rm A}}-c_{{\rm A}}^{2}\Ident\equiv-c_{{\rm A}}^{2}\Pperp$.

To represent $\vsf{D}$ as a matrix we project onto the orthonormal
basis \ref{eq:waveBasis}, which can be written $\esub{1}=\vec{k}_{\perp}/k_{\perp}\;,\;\esub{2}=\vec{c}_{{\rm A}}/c_{{\rm A}}\;\;\text{and}\;\;\esub{3}=\vec{k}_{\perp}\cross\vec{c}_{{\rm A}}/\left(k_{\perp}c_{{\rm A}}\right)\;.$
We thus have 
\begin{align}
\vsf{D} & =\left(\omega_{\vec{k}}^{\prime\vec{u}}+\epsRx k_{\parallel}u^{{\rm Rx}}\right)\Ident-\epsRx u^{{\rm Rx}}\left(k_{\perp}\esub{1}+k_{\parallel}\esub{2}\right)\esub{2}-\epsI\frac{c_{{\rm A}}^{2}}{\omega_{\vec{k}}^{\prime\vec{u}}}\left(k^{2}\esub{1}\esub{1}+k_{\parallel}^{2}\esub{3}\esub{3}\right)\nonumber \\
 & \qquad\qquad\qquad\qquad\qquad\qquad-\frac{c_{{\rm s}}^{2}}{\omega_{\vec{k}}^{\prime\vec{u}}}\left(k_{\perp}\esub{1}+k_{\parallel}\esub{2}\right)\left(k_{\perp}\esub{1}+k_{\parallel}\esub{2}\right)\;,\label{eq:Ddyads}
\end{align}
which can be represented as the block-diagonal matrix

\begin{equation}
\vsf{D}=\begin{bmatrix}\vsf{D}_{{\rm MS}} & 0\\
0 & \omega_{\vec{k}}^{\prime\vec{u}}+\epsRx k_{\parallel}u^{{\rm Rx}}-\epsI k_{\parallel}^{2}c_{{\rm A}}^{2}/\omega_{\vec{k}}^{\prime\vec{u}}
\end{bmatrix}\;,\label{eq:Dblocks}
\end{equation}
with the $1\times1$ \emph{Alfvén block} on the lower right and the
$2\times2$ \emph{magnetosonic block}, 
\begin{equation}
\vsf{D}_{{\rm MS}}=\begin{bmatrix}\omega_{\vec{k}}^{\prime\vec{u}}+\epsRx k_{\parallel}u^{{\rm Rx}}-\left(\epsI k^{2}c_{{\rm A}}^{2}+k_{\perp}^{2}c_{{\rm s}}^{2}\right)/\omega_{\vec{k}}^{\prime\vec{u}} & -k_{\perp}\left(\epsRx u^{{\rm Rx}}+k_{\parallel}c_{{\rm s}}^{2}/\omega_{\vec{k}}^{\prime\vec{u}}\right)\\
-k_{\parallel}k_{\perp}c_{{\rm s}}^{2}/\omega_{\vec{k}}^{\prime\vec{u}} & \omega_{\vec{k}}^{\prime\vec{u}}-k_{\parallel}^{2}c_{{\rm s}}^{2}/\omega_{\vec{k}}^{\prime\vec{u}}
\end{bmatrix}\;,\label{eq:DMS}
\end{equation}
upper left.

\subsubsection{Limiting cases}

Consider first the ideal, fully converged case $\cIOL=0$ ($\epsI=1,\;\epsRx=0$)
and use \eqref{eq:LemXIX} to write $k^{2}\Ident-\vec{k}\vec{k}=-\vec{k}\cross\Ident\cross\vec{k}$
so

\begin{align*}
\vsf{D} & =\omega_{\vec{k}}^{\prime\vec{u}}\Ident-\frac{\vec{B}\cross\left(\vec{k}\cross\Ident\cross\vec{k}\right)\cross\vec{B}}{\muSI\rho\,\omega_{\vec{k}}^{\prime\vec{u}}}-\text{\ensuremath{\tau_{\Omega}}}\frac{\vec{k}\vec{k}}{\omega_{\vec{k}}^{\prime\vec{u}}}\\
 & =\omega_{\vec{k}}^{\prime\vec{u}}\Ident-\frac{\left(\vec{k}\vec{B}-\vec{k}\dotv\vec{B}\Ident\right)\dotv\left(\vec{B}\vec{k}-\vec{k}\dotv\vec{B}\Ident\right)}{\muSI\rho\,\omega_{\vec{k}}^{\prime\vec{u}}}-\text{\ensuremath{\tau_{\Omega}}}\frac{\vec{k}\vec{k}}{\omega_{\vec{k}}^{\prime\vec{u}}}\;,
\end{align*}
which, apart from the definitions of $\vsf{D}$ differing by a factor
of $\rho\,\omega_{\vec{k}}^{\prime\vec{u}}$, agrees with the IMHD
form, eq. (75), of \citet{Dewar_Burby_Qu_Sato_Hole_2020}.

In the pure RxMHD case $\epsI=0,\;\epsRx=1$, 
\[
\vsf{D}=\omega_{\vec{k}}^{\prime\vec{u}}\Ident-\left(\vec{k}\uRx-\vec{k}\dotv\uRx\Ident\right)-\text{\ensuremath{\tau_{\Omega}}}\frac{\vec{k}\vec{k}}{\omega_{\vec{k}}^{\prime\vec{u}}}\;.
\]
Apart from the definitions of $\vsf{D}$ again differing by a factor
of $\rho\,\omega_{\vec{k}}^{\prime\vec{u}}$, this agrees with the
RxMHD form, eq. (88), of \citet{Dewar_Burby_Qu_Sato_Hole_2020}.

Thus $\epsRx$ parametrizes a continuous interpolation between RxMHD
and IMHD.

\begin{figure}
\includegraphics{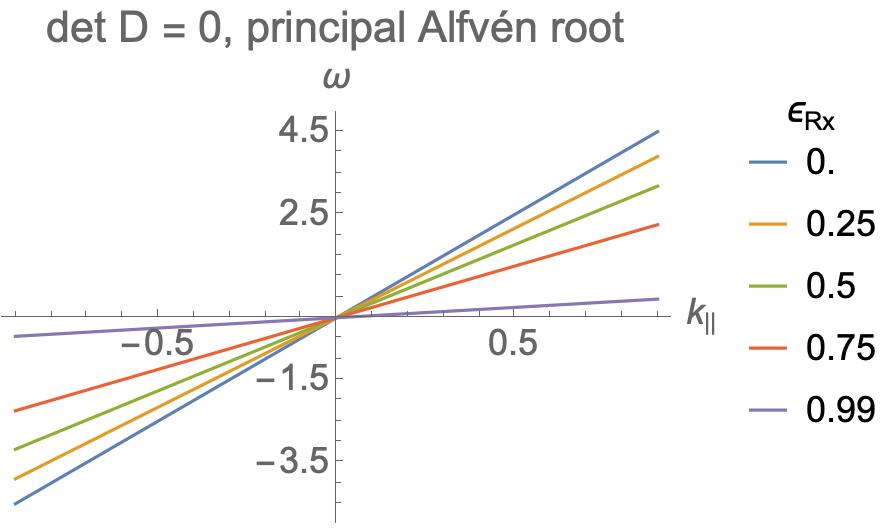}

\caption{Showing transition of the Alfvén-branch dispersion relation: $\omega$
vs. $k_{\parallel}$; IMHD ($\epsRx=0$) to RxMHD ($\epsRx=1$: We
have used $\epsRx=0.99$ for clarity as the $\epsRx=1$ line coincides
with the $k_{\parallel}$ axis). Fixed parameters (in arb. units)
are $k=1$, $\vec{u}=0$, $\uRx=0$, $c_{{\rm s}}=1$ and $c_{{\rm A}}=5$.
(Colour online. The vertical ordering of the lines in the $k_{\parallel}>0$
half plane coincides with that of $\epsRx$ in the legend.)\label{fig:Alfven}}
\end{figure}

\subsubsection{Dispersion relations --- Alfvén branches}

Multiplying the first factor of the determinant 
\begin{equation}
\det\vsf{D}=\left(\omega_{\vec{k}}^{\prime\vec{u}}+\epsRx k_{\parallel}u^{{\rm Rx}}-\epsI k_{\parallel}^{2}c_{{\rm A}}^{2}/\omega_{\vec{k}}^{\prime\vec{u}}\right)\det\vsf{D}_{{\rm MS}}\label{eq:detD}
\end{equation}
by $\omega_{\vec{k}}^{\prime\vec{u}}$ gives the dispersion relation
for the Alfvén-wave branch(es) as the quadratic equation 
\begin{equation}
\left(\omega_{\vec{k}}^{\prime\vec{u}}\right)^{2}+\epsRx k_{\parallel}u^{{\rm Rx}}\omega_{\vec{k}}^{\prime\vec{u}}-\left(1-\epsRx\right)k_{\parallel}^{2}c_{{\rm A}}^{2}=0\;.\label{eq:AlfvenDisprel}
\end{equation}
The general solution of the quadratic equation is 
\begin{equation}
\omega_{\vec{k}}^{\prime\vec{u}}=\frac{k_{\parallel}}{2}\left[-\epsRx u^{{\rm Rx}}\pm\left(4\epsI c_{{\rm A}}^{2}+\left(\epsRx u^{{\rm Rx}}\right)^{2}\right)^{1/2}\right]\label{eq:AlfvenDispSoln}
\end{equation}
Qualitative analysis is more informative: As $\epsRx\to0$ the dispersion
relations for the two branches approach the Doppler-shifted Alfvén-wave
dispersion relations $\omega-\vec{k}\dotv\vec{u}=\pm k_{\parallel}c_{{\rm A}}$.
Also, inspection shows that $\omega-\vec{k}\dotv\vec{u}\to0$ as $k_{\parallel}\to0$
quite generally, and when $\left|\epsRx\right|\ll1$ the modification
of the dispersion departure from the standard Alfvén-wave dispersion
relation is essentially determined by the \emph{product} $\epsRx u^{{\rm Rx}}$.
Thus, when when $\left|\epsRx\right|\ll1$ and the parallel flow parameter
is at most Alfvénic, $u^{{\rm Rx}}/c_{{\rm A}}\leq O(1)$, $\epsRx$
and $u^{{\rm Rx}}$ will have little effect on the Alfvén-wave branches.

The plots in figure \ref{fig:Alfven} give a visualization of the
dependence of the dispersion relation on $\epsRx$. The figure is
for a case where $\uRx=0$, when \eqref{eq:AlfvenDispSoln} simplifies
to $\omega_{\vec{k}}^{\prime\vec{u}}=\pm\sqrt{\epsI}\,k_{\parallel}c_{{\rm A}}$.
(We call the $+$ solution the \emph{principal branch}.) The square
root term $\sqrt{\epsI}=\sqrt{1-\epsRx}$ gives rise to a singular
dependence on $\epsRx$ at $\epsRx=1$ but the vicinity of IMHD is
regular.

\begin{figure}
\includegraphics{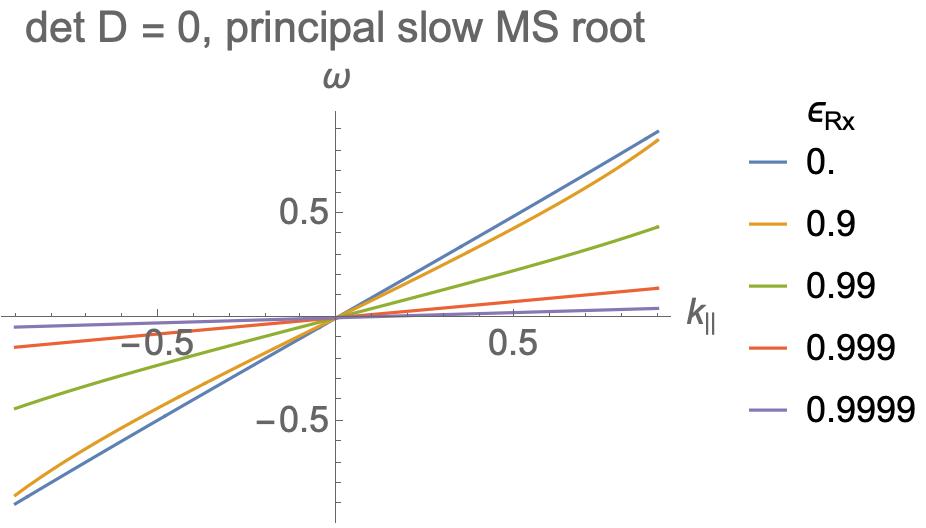}

\caption{Showing transition of the slow-magnetosonic-branch dispersion relation:
$\omega$ vs. $k_{\parallel}$; from IMHD ($\epsRx=0$) to RxMHD ($\epsRx=1$).
Fixed parameters (in arb. units) are $k=1$, $\vec{u}=0$, $\uRx=0$,
$c_{{\rm s}}=1$ and $c_{{\rm A}}=5$. (Colour online. The vertical
ordering of the lines in the $k_{\parallel}>0$ half plane coincides
with that of $\epsRx$ in the legend.)\label{fig:slowMS}}
\end{figure}

\begin{figure}
\includegraphics{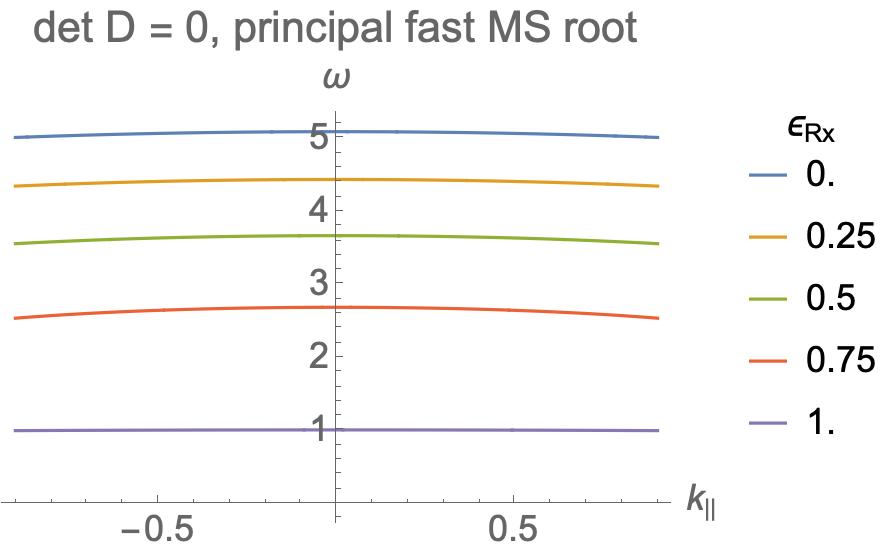}\caption{Showing transition of the fast-magnetosonic-branch dispersion relation:
$\omega$ vs. $k_{\parallel}$; from IMHD ($\epsRx=0$) to RxMHD ($\epsRx=1$).
Fixed parameters (in arb. units) are $k=1$, $\vec{u}=0$, $\uRx=0$,
$c_{{\rm s}}=1$ and $c_{{\rm A}}=5$. (Colour online. The vertical
ordering of the lines in the $k_{\parallel}>0$ half plane coincides
with that of $\epsRx$ in the legend.)\label{fig:fastMS}}
\end{figure}

\subsubsection{Dispersion relations --- Magnetosonic branches}

The magnetosonic dispersion relations are obtained by setting $\det\vsf{D}_{{\rm MS}}=0$,
where

\[
\det\vsf{D}_{{\rm MS}}=\frac{c_{{\rm A}}^{2}\epsI k^{2}\left(k_{\parallel}^{2}c_{{\rm s}}^{2}-\omega^{2}\right)-\omega\left(c_{{\rm s}}^{2}k^{2}-\omega^{2}\right)(\epsRx k_{\parallel}u^{{\rm Rx}}+\omega)}{\omega^{2}}\;.
\]

The solution of the quartic equation $\omega^{2}\det\vsf{D}_{{\rm MS}}=0$
is extremely complicated but the figures \ref{fig:slowMS} and \ref{fig:fastMS}
give an overview of the $\epsRx$ dependence. Again, the limit $\epsRx\to1$
is clearly singular in the slow magnetosonic case but not $\epsRx\to0$.
This regularity around ideal MHD means our dispersion relation analysis
is too crude to reveal the potential regularizing effect of softening
the IOL constraint.

\section{Conclusion\label{sec:Concl}}

Invoking the augmented Lagrangian version of the penalty function
method for constrained optimization, we have sketched out what we
hope is a practical computational approach for iteratively solving
the Relaxed MHD (RxMHD) Euler--Lagrange equations of \citet{Dewar_Burby_Qu_Sato_Hole_2020}with
added Ideal Ohm's Law (IOL) constraint terms.

This method depends crucially on the existence of a Lagrange multiplier
field to be found using the augmented Lagrangian iteration algorithm
borrowed from finite-dimensional optimization theory.

A formal proof of convergence in may in general be difficult, but
a practical approach will be to test the algorithm by perturbing away
from IOL-feasible relaxed equilibria in simple geometries. A suitable
such starting point is the rigidly rotating axisymmetric tokamak equilibrium
discussed by \citet{Dewar_Burby_Qu_Sato_Hole_2020}. In this paper
have illustrated the construction of the Lagrange multiplier field
for linearized wave perturbations in the short-wavelength WKB approximation.

To find the constrained momentum equation we have used a little-known
dyadic identity to derive a general conservation form. Substituting
the constrained-RxMHD Lagrangian into this general form reveals residual
terms in the stress tensor and a fictitious external force that should
tend to zero uniformly in $\Omega$ if the constraint iteration converges
so as to satisfy the IOL equality constraint.

However in non-axisymmetric, three-dimensional (3-D) plasma confinement
systems such as stellarators and real tokamaks with field errors and
intentionally resonant magnetic perturbations, there is good physical
reason to believe uniform pointwise convergence is impossible. In
such cases the best we can hope for is convergence in an $L^{2}$-norm,
which will provide a weak-form regularization to cope with the singularities
to which IMHD is prone in 3-D. This regularization should break the
frozen-in flux condition of IMHD on small scales and allow interesting
behaviour to be simulated without raising the order of the PDEs as
adding resistivity does. Potential applications include reconnection
events and the conjectured formation of equilibrium fractal magnetic
and fluid flow patterns in 3-D systems. Other potential physical phenomena
to investigate in 3-D systems include the linear normal mode spectrum,
nonlinear saturation, bifurcations to oscillatory modes, and the effect
of quasisymmetry {[}\citet{Nuehrenberg_Zille_88}; \citet{Burby_Kallinikos_MacKay_2020};
\citet{Rodriguez_Helander_Bhattacharjee_2020}; \citet{Constantin_Drivas_Ginsberg_2021}{]}
on 3-D equilibria with flow {[}\citet{Vanneste_Wirosoetisno_2008}{]}.

Also, to improve the physical applicability of relaxed MHD it will
be important to extend the handling of thermal relaxation beyond isotropic
pressure. Relaxation parallel to the magnetic field is very reasonable
physically but perpendicular relaxation has forced the use of discontinuous
pressure profiles in the MRxMHD-based SPEC code described by \citet{Hudson_etal_2012b}.
Thus it will be important to build on the work of \citet{Dennis_Hudson_Dewar_Hole_2014b}
to include an anisotropic pressure tensor in a weakly IOL-feasible
model.

N.B. An unabridged version of this paper with more detail on derivations
of equations is available online as Supplementary Material at <link
to be inserted by editors>.

\section*{Appendices}

\appendix

\section{A very brief history of relaxed MHD\label{sec:RxHistory}}

The term \emph{relaxation} in the physical sciences generally connotes
a process by which a system tends toward an equilibrium state: thermodynamic,
chemical, electrodynamic, mechanical, or a combination of these. For
example, in a closed, constant energy system initially out of thermodynamic
equilibrium, relaxation occurs as the entropy increases toward a \emph{maximum}.
In an open system at a temperature above that of a surrounding heat
bath, relaxation occurs as heat carries energy out of the system,
so its thermal energy tends toward a \emph{minimum}.

In an open system with unbalanced mechanical forces, potential energy
is converted into kinetic energy, which in turn is dissipated by friction
into heat that is lost to the outside world, thus minimizing \emph{total
}energy, thermal and potential. This is the paradigm implicit in our
use of the term ``relaxation'', the assumption that a relaxed state
is defined by the minimum of a Hamiltonian.

In plasma physics the first use of the term may have been in the paper
by \citet{Chandrasekhar_Woltjer_1958}, which proposes two variational
principles other than maximizing entropy or minimizing energy: \emph{maximum
energy for given mean-square current density} and \emph{minimum dissipation
for a given magnetic energy. }The common element in these, and the
\emph{minimum energy at constant magnetic helicity} principle used
by \citet{Woltjer_58a} and Taylor \citet{Taylor_74} is the derivation
of a ``linear-force-free'' magnetic field obeying the Beltrami equation
$\curl\vec{B}=\mu\vec{B}$, with $\mu$ constant, as the outcome.
The Chandrasekhar and Woltjer work was in the context of plasma astrophysics,
justifying the force-free assumption (where the force density in question
is $\vec{j}\cross\vec{B}$) basically on the assumption the plasma
has low $\beta\defeq p/\left(B^{2}/2\muSI\right)$ and no confining
forces that are strong compared with gradients of magnetic pressure.
In contrast Taylor considered a toroidal terrestrial plasma confined
in a metal shell and driven by a strong induced current, creating
a turbulent state from which the plasma relaxes. Taylor regards the
relaxation mechanism as the breaking of the microscopic IMHD topological
invariants leaving only the global magnetic helicity as conserved.

\citet{Taylor_74} is uncommital as to the exact details of this breaking
of microscopic invariants and is content to use successful comparison
with experiment of the predictions flowing from his derivation of
the Beltrami equation as sufficient validation of his elegantly simple
model, a general philosophy we also adopt. However in his later review,
\citet{Taylor_86} gives some more detail on the decay mechanism,
citing some turbulence simulations and invokes turbulence scale length
arguments to explain why it is energy that is minimized rather than
magnetic helicity. \citet{Moffatt_2015} has recently critically reviewed
the arguments of \citet{Taylor_86} from a more modern perspective.

\citet{Woltjer_58b} pointed out there were other global IMHD invariants
beyond magnetic helicity, in particular his eq. (2), the cross helicity
involving both flow and magnetic field. \citet{Bhattacharjee_Dewar_82}
pointed that in an axisymmetric system an infinity of additional global
invariants could be generated by taking moments of $\vec{A}\dotv\vec{B}$
with powers of a flux function, and used lower moments to generate
more physical pressure and current profiles for tokamak equilibria
than the very restricted profiles given by Taylor's relaxation principle.
\citet{Hudson_etal_2012b} developed multi-region relaxed MHD (MRxMHD),
a generalization of single-region Taylor relaxation by inserting thin
IMHD barrier interface tori to frustrate global Taylor relaxation.
This generalization is appropriate to non-axisymmetric equilibria
in stellarators and in tokamaks with symmetry-breaking perturbations,
where magnetic field-line flow can be chaotic even without turbulence.
This MRxMHD formulation is implemented in the now well-established
Stepped-Pressure Equilibrium Code (SPEC).

A relaxation approach for finding equilibria with \emph{flow} by adding
a constraint additional to conservation of magnetic helicity, conservation
of cross helicity, was used by Finn and Antonsen \citet{Finn_Antonsen_83}
using an entropy-maximization relaxation principle {[}see also the
contemporaneous paper by Hameiri \citet{Hameiri_83}{]}. However,
they show this leads to the same equations as energy minimization.
Thus we take, as in IMHD, the entropy in $\Omega$ to be conserved
and follow Taylor in defining relaxed states as energy minima.

Pseudo-dynamical energy-descent relaxation processes that conserve
topological invariants have been developed, \citet{Vallis_Carnevale_Young_89,Vladimirov_Moffatt_Ilin_99}
but we stay within the framework of conservative classical mechanics
by developing a dynamical formalism, RxMHD, that includes relaxed
equilibria as stationary points of a relaxation Hamiltonian, with
Lagrange multipliers to constrain chosen macroscopic invariants, but
which also allows non-equilibrium motions, most easily done using
Hamilton's action principle. Stability can also be examined by taking
the second variation of the Hamiltonian, \citet{Vladimirov_Moffatt_Ilin_99}
but in this paper, as in \citet{Dewar_Yoshida_Bhattacharjee_Hudson_2015}
and \citet{Dewar_Burby_Qu_Sato_Hole_2020}we deal only with first
variations.

However the SPEC code implements a Newton method for finding energy
minima and saddle points by calculating a Hessian matrix, which is
the second variation of the MRxMHD energy. Combined with a model kinetic
energy obtained by loading all mass onto the interfaces between the
relaxation regions. This has recently been used successfully by \citet{Kumar_Qu_etal_2021,Kumar_etal_2021}
for calculating the spectrum of some linear eigenmodes in a tokamak,
but comparison between the model kinetic energy and our new dynamical
relaxation theory is desirable for determining the domain of applicability
of the mass loading model.

\section{Some vector and dyadic identities\label{sec:lemmas}}

In the body of this paper we have used the usual coordinate-free vector
(and dyadic) calculus notations, but in this appendix we derive some
identities that are more easily proved using elementary tensor notation.
Assuming an arbitrary fixed orthonormal basis $\{\esub{i}\}$, $i=1,2,3\mod3$,
a vector, $\vec{a}$ say, is represented as $\vec{a}=a_{i}\esub{i}$,
the summation convention for contraction over repeated dummy indices
being assumed throughout.

Thus dot and cross products are represented as $\vec{a}\cdot\vec{b}=a_{i}b_{i}$
and $\vec{a}\cross\vec{b}=\esub{i}\varepsilon_{i,j,k}a_{j}b_{k}$,
respectively, where the alternating Levi-Civita tensor $\varepsilon_{ijk}$
is $1$ or $-1$ according as $\{i,j,k\}$ is an even or odd permutation
of $\{1,2,3\}$, or $0$ if it is neither (e.g. if there are repeated
integers). Also the operations of grad and curl acting on scalar and
vector functions $f$ and $\vec{f}$, respectively, are represented
as $\grad f=\partial f/\partial\vec{x}\defeq\esub{i}\partial_{i}f$
and $\curl\vec{a}\defeq\esub{i}\varepsilon_{ijk}\partial_{j}a_{k}$,
where $\partial_{i}\defeq\partial/\partial x_{i}$. We use parentheses
to limit the scope of the rightward differentiation of such operators.
NB Left-right ordering is more important in vector notation. E.g.
the dyadics $\vec{a}\vec{b}$ and $\vec{b}\vec{a}$ are distinct,
but $a_{i}b_{j}=b_{j}a_{i}$.

First we derive three useful identities involving gradients with respect
to $\vec{B}\equiv B_{i}\esub{i}$, and the unit vector parallel to
$\vec{\ensuremath{B}}$, $\esub{\vec{B}}(\vec{x})\equiv\vec{B}(\vec{x})/B(\vec{x})$.
(By ``parallel to $\vec{\ensuremath{B}}$'' we mean locally tangent
to the magnetic field line passing though any point $\vec{x}$. Henceforth
the dependence on $\vec{x}$ is implicit as these identities concern
functions purely of $\vec{B}$.):

\begin{lemma}

The gradients of $\vec{B}$, $B$ and $\esub B$ with respect to $\vec{B}$
are, in terms of the identity dyadic $\Ident\defeq\sum_{i}\esub{i}\esub{i}$,
the unit tangent vector $\esub{\vec{B}}$, and $\Pperp\defeq\Ident-\esub{\vec{B}}\esub{\vec{B}}$,
the projector onto the plane perpendicular to $\vec{B}$,

\begin{equation}
\frac{\partial\vec{B}}{\partial\vec{B}}=\Ident,\quad\frac{\partial B}{\partial\vec{B}}=\esub{\vec{B}},\;\;\text{and}\quad\frac{\partial\esub{\vec{B}}}{\partial\vec{B}}=\frac{\Pperp}{B}\;.\label{eq:Bidents}
\end{equation}

\end{lemma}

\emph{Derivations: }Using the notations $\partial_{\vec{B}}\cdot\equiv\esub{i}\partial_{B_{i}}\equiv\partial\cdot/\partial\vec{B}$,
we have the obvious identity $\partial_{\vec{B}}\vec{B}=\Ident$.
Applying this first identity to $B\equiv(\vec{B}\dotv\vec{B})^{1/2}$
we find the second identity, $\partial_{\vec{B}}B=\left(2\Ident\dotv\vec{B}\right)/2B=\vec{B}/B=\esub{\vec{B}}$.
The third identity follows from the first two: $\partial_{\vec{B}}\left(\vec{B}/B\right)=\Ident/B-\vec{B}$$\esub{\vec{B}}$$/B^{2}=\left(\Ident-\esub{\vec{B}}\esub{\vec{B}}\right)/B$.
$\quad\Box$\\

\begin{lemma} Variational derivative of functional $F[\vec{A},\Phi]=\iint\!\!f(\vec{A},\vec{B},\vec{E})\d V\d t$
is 
\begin{equation}
\frac{\delta F}{\delta\vec{A}}=\frac{\partial f}{\partial\vec{A}}+\curl\frac{\partial f}{\partial\vec{B}}+\epsF\frac{\partial}{\partial t}\frac{\partial f}{\partial\vec{E}}\;,\label{eq:LemdFdA}
\end{equation}
where $f$ is an arbitrary scalar-valued function of $\vec{A}$, $\vec{B}=\curl{\vec{A}}$,
and $\vec{E}=-\epsF\partial_{t}$$\vec{A}-\grad\Phi$ from \eqref{eq:Erep},
$\vec{A}$ being an arbitrary vector field. \end{lemma}

Varying $\vec{A}$ 
\begin{align*}
\delta F & =\iint\left[\frac{\partial f}{\partial\vec{A}}\dotv\delta\vec{A}+\frac{\partial f}{\partial\vec{B}}\dotv\curl\delta\vec{A}-\epsF\frac{\partial f}{\partial\vec{E}}\dotv\partial_{t}\delta\vec{A}\right]\d V\d t\\
 & =\iint\left[\frac{\partial f}{\partial A_{i}}\delta A_{i}+\frac{\partial f}{\partial B_{i}}\varepsilon_{i,j,k}\partial_{j}\delta A_{k}-\epsF\frac{\partial f}{\partial E_{i}}\partial_{t}\delta A_{i}\right]\d V\d t\\
 & =\iint\left[\frac{\partial f}{\partial A_{i}}\delta A_{i}-\varepsilon_{k,j,i}\left(\partial_{j}\frac{\partial f}{\partial B_{k}}\right)\delta A_{i}+\epsF\left(\partial_{t}\frac{\partial f}{\partial E_{i}}\right)\delta A_{i}\right]\d V\d t\;,\quad i\rightleftarrows k\:\&\:\text{ibp}\\
 & =\iint\left[\frac{\partial f}{\partial A_{i}}+\varepsilon_{i,j,k}\left(\partial_{j}\frac{\partial f}{\partial B_{k}}\right)+\epsF\left(\partial_{t}\frac{\partial f}{\partial E_{i}}\right)\right]\delta A_{i}\,\d V\d t\;,\quad\varepsilon_{k,j,i}=-\varepsilon_{i,j,k}\\
 & =\iint\left(\frac{\partial f}{\partial\vec{A}}+\curl\frac{\partial f}{\partial\vec{B}}+\epsF\frac{\partial}{\partial t}\frac{\partial f}{\partial\vec{E}}\right)\!\dotv\!\delta\vec{A}\,\d V\d t\defeq\int\!\frac{\delta F}{\delta\vec{A}}\dotv\delta\vec{A}\,\d V\d t\;,
\end{align*}
where $\rightleftarrows$ stands for ``have swapped dummy indices''
and ``ibp'' stands for ``have integrated by parts'' (neglecting
surface terms on the assumption that the supports of variations do
not include the boundary).$\quad\Box$

\begin{lemma} Variational derivative of functional $F$ above is

\begin{equation}
\frac{\delta F}{\delta\Phi}=\divv\frac{\partial f}{\partial\vec{E}}\;,\label{eq:LemdFdPhi}
\end{equation}

\end{lemma}

\emph{Derivation: }Varying \emph{$\Phi$}

\begin{align*}
\delta F & =\iint\left[-\frac{\partial f}{\partial\vec{E}}\dotv\grad\delta\Phi\right]\d V\d t\\
 & =\iint\left[\left(\divv\frac{\partial f}{\partial\vec{E}}\right)\delta\Phi\right]\d V\d t\defeq\int\!\frac{\delta F}{\delta\Phi}\delta\Phi\,\d V\d t\;,
\end{align*}
neglecting surface term as above.$\quad\Box$

Two useful identities, closely related to integration by parts, for
deriving conservation forms of Euler--Lagrange equations for freely
variable fields {[}members of the set denoted $\bm{\upeta}$ in \citet{Dewar_Burby_Qu_Sato_Hole_2020}{]}
are

\begin{lemma} For scalar fields, e.g. $\Phi$ 
\begin{equation}
(\grad\grad\Phi)\dotv\vec{f}=\divv[\vec{f}\grad\Phi]-(\grad\Phi)\divv\vec{f}\;,\label{eq:LemgradgradPhi}
\end{equation}
where $\vec{f}$ is an arbitrary vector field, e.g. $\partial\mathcal{L}/\partial\grad\Phi$.

\end{lemma}

\emph{Derivation: }Follows directly from fact $\grad\grad\Phi$ is
a symmetric dyadic, proved in first line below, 
\begin{align*}
(\grad\grad\Phi)\dotv\vec{f} & =\esub{i}(\partial_{i}\partial_{j}\Phi)f_{j}=\esub{i}(\partial_{j}\partial_{i}\Phi)f_{j}\;,\quad\partial_{i}\rightleftharpoons\partial_{j}\\
 & =\esub{i}(\partial_{j}\partial_{i}\Phi f_{j})-\esub{i}(\partial_{i}\Phi)\partial_{j}f_{j}\\
 & =\divv[\vec{f}\grad\Phi]-(\grad\Phi)\divv\vec{f}\;,
\end{align*}
where $\rightleftharpoons$ stands for ``have commuted operators''.$\Box$

\begin{corollary} For $\vec{E}=-\grad\Phi-\epsF\partial_{t}\vec{A}$,
\begin{equation}
\left(\grad\vec{E}\right)\dotv\vec{f}=\divv[\vec{f}\vec{E}]-\vec{E}\divv\vec{f}-\epsF\vec{f}\cross\partial_{t}\vec{B}\;.\label{eq:LemgradEdotf}
\end{equation}

\end{corollary}

\emph{Derivation: }Muliplying each side of \eqref{eq:LemgradgradPhi}
by $-1$, writing $-\grad\Phi=\vec{E}+\epsF\partial_{t}\vec{A}$ and
subtracting $\epsF\partial_{t}\vec{A}$ from both sides, the lemma
\eqref{eq:LemgradgradPhi} becomes

\begin{align*}
\left(\grad\vec{E}\right)\dotv\vec{f} & =\divv[\vec{f}\vec{(E}+\epsF\partial_{t}\vec{A})]-\vec{(E}+\epsF\partial_{t}\vec{A})\divv\vec{f}-\epsF(\grad\partial_{t}\vec{A})\dotv\vec{f}\;,\\
 & =\divv[\vec{f}\vec{E}]-\vec{E}\divv\vec{f}+\epsF\divv[\vec{f}\partial_{t}\vec{A}]-\epsF\left(\partial_{t}\vec{A}\right)\divv\vec{f}-\epsF(\grad\partial_{t}\vec{A})\dotv\vec{f}\\
 & =\divv[\vec{f}\vec{E}]-\vec{E}\divv\vec{f}+\epsF\vec{f}\dotv(\grad\partial_{t}\vec{A})-\epsF(\grad\partial_{t}\vec{A})\dotv\vec{f}\\
 & =\divv[\vec{f}\vec{E}]-\vec{E}\divv\vec{f}-\epsF\vec{f}\cross(\curl\partial_{t}\vec{A})\\
 & =\divv[\vec{f}\vec{E}]-\vec{E}\divv\vec{f}-\epsF\vec{f}\cross\partial_{t}\vec{B}\;.\Box
\end{align*}

\begin{lemma} For $\vec{A}$ an arbitrary vector field and $\vec{B}=\curl\vec{A},$
\begin{equation}
(\grad\vec{B})\dotv\vec{f}=-\divv[\vec{f}\cross(\grad\vec{A})^{{\rm T}}]+(\grad\vec{A})\dotv\curl\vec{f}\;,\label{eq:Lem3}
\end{equation}
where $\vec{f}$ is an arbitrary vector field, e.g. $\partial\mathcal{L}/\partial\vec{B}$.
{[}N.B. For a more useful form see the corollary \eqref{eq:LemgradBdotf}
below.{]}

\end{lemma}

\emph{Derivation: } 
\begin{align*}
(\grad\curl\vec{A})\dotv\vec{f} & =\esub{i}(\partial_{i}\varepsilon_{j,k,l}\partial_{k}A_{l})f_{j}=\esub{i}(\partial_{k}\varepsilon_{j,k,l}\partial_{i}A_{l})f_{j}\;,\quad\partial_{i}\rightleftharpoons\partial_{k}\\
 & =\esub{i}\partial_{k}[(\partial_{i}A_{l})\varepsilon_{j,k,l}f_{j}]-\esub{i}(\partial_{i}A_{l})\varepsilon_{j,k,l}\partial_{k}f_{j}\\
 & =-\partial_{k}[\varepsilon_{k,j,l}f_{j}(\grad A_{l})]+(\grad A_{l})\varepsilon_{l,k,j}\partial_{k}f_{j}\;,\quad\text{anticyclic perms. of }j,k,l\\
 & =-\divv[\vec{f}\cross(\grad\vec{A})^{{\rm T}}]+(\grad\vec{A})\dotv\curl\vec{f}\;,
\end{align*}
where $(\grad\vec{A})^{{\rm T}}$ is the transpose of the dyadic $\grad\vec{A}$.$\quad\Box$

\begin{corollary}

\begin{align}
(\grad\vec{B})\dotv\vec{f} & =\left(\curl\vec{f}\right)\cross\vec{B}-\divv\left[\vec{f}\cross\Ident\cross\vec{B}\right]\label{eq:LemgradBdotf}
\end{align}

\end{corollary}

\emph{Derivation:} Because of the identity $\divv\left(\vec{f}\cross\grad\vec{A}\right)-\left(\curl\vec{f}\right)\dotv\grad\vec{A}=0$
(which is easily proven using the properties of the scalar product
and the identity $\curl\grad=0$) we can add $\divv\left(\vec{f}\cross\grad\vec{A}\right)-(\grad\vec{A})^{{\rm T}}\dotv\left(\curl\vec{f}\right)$
to the RHS of \eqref{eq:Lem3} to antisymmetrize $\grad\vec{A}$ and
thus to eliminate it in favour of $\curl\vec{A}=\vec{B}$:

\begin{align*}
(\grad\vec{B})\dotv\vec{f} & =\divv\left[\vec{f}\cross\left(\grad\vec{A}-(\grad\vec{A})^{{\rm T}}\right)\right]+\left(\grad\vec{A}-(\grad\vec{A})^{{\rm T}}\right)\dotv\curl\vec{f}\\
 & =-\divv\left[\vec{f}\cross\Ident\cross\left(\curl\vec{A}\right)\right]+\left(\curl\vec{f}\right)\cross\left(\curl\vec{A}\right)
\end{align*}
the second term in the second line following from

\begin{align*}
\vec{f}\cross\left(\grad\vec{A}-(\grad\vec{A})^{{\rm T}}\right) & =\vec{f}\cross\esub{i}\esub{i}\dotv\left(\grad\vec{A}-(\grad\vec{A})^{{\rm T}}\right)\\
 & =-\vec{f}\cross\esub{i}\esub{i}\cross\left(\curl\vec{A}\right)\quad\Box
\end{align*}
Alternatively, \emph{verify} without using vector potential but assuming
$\divv\vec{B}=0$:

\begin{align*}
\text{RHS} & =\left(\curl\vec{f}\right)\cross\vec{B}-\divv\left[\vec{f}\cross\esub i\esub i\cross\vec{B}\right]\\
 & =\:\left(\curl\vec{f}\right)\cross\vec{B}-\left(\curl\vec{f}\right)\dotv\esub i\esub i\cross\vec{B}\\
 & \qquad+\vec{f}\dotv\esub j\cross\grad\vec{B}\cross\esub j\\
 & =-\esub mf\varepsilon_{j,i,k}\varepsilon_{j,m,l}\partial_{k}B_{l}\\
 & =-\esub mf_{i}\left(\delta_{i,m}\delta_{k,l}-\delta_{i,l}\delta_{k,m}\right)\partial_{k}B_{l}\\
 & =\esub kf_{i}\partial_{k}B_{i}-\esub if_{i}\partial_{k}B_{k}\\
 & =\left(\grad\vec{B}\right)\dotv\vec{f}-\vec{f}\divv\vec{B}\\
 & =\text{LHS}\quad\Box\;,
\end{align*}
where we used the Levi-Civita tensor contraction result $\varepsilon_{a,b,c}\varepsilon_{a,j,k}=\delta_{b,j}\delta_{c,k}-\delta_{b,k}\delta_{c,j}$,
where here $\delta$ is the Kronecker symbol.\footnote{This contraction result was obtained using the helpful tool at \url{https://demonstrations.wolfram.com/ProductOfTwoLeviCivitaTensorsWithContractions/}
.}

\begin{lemma}

\begin{equation}
\vec{f}\cross\Ident\cross\vec{g}=\vec{g}\vec{f}-\vec{{f}}\!\dotv\!\vec{{g}}\,\Ident\label{eq:LemXIX}
\end{equation}

\end{lemma}

\emph{Derivation:} 
\begin{align*}
\left(\vec{f}\cross\Ident\cross\vec{g}\right)_{i,j} & =f_{k}\esub{i}\dotv\esub{k}\cross\esub{l}\esub{l}\cross\esub{m}\dotv\esub{j}g_{m}\\
 & =f_{k}\varepsilon_{i,k,l}\varepsilon_{l,m,j}g_{m}=f_{k}\varepsilon_{l,i,k}\varepsilon_{l,m,j}g_{m}\\
 & =f_{k}\left(\delta_{i,m}\delta_{k,j}-\delta_{i,j}\delta_{k,m}\right)g_{m}\\
 & =f_{j}g_{i}-f_{k}g_{k}\delta_{i,j}\:\:\Box
\end{align*}

\emph{Verification:}

\begin{align*}
\vec{a}\dotv\text{\text{LHS}} & =\vec{a}\dotv\vec{f}\cross\Ident\cross\vec{g}\\
 & =\vec{a}\cross\vec{f}\dotv\Ident\cross\vec{g}\\
 & =\left(\vec{a}\cross\vec{f}\right)\cross\vec{g}\\
 & =\left(\vec{a}\dotv\vec{g}\right)\vec{f}-\left(\vec{f}\dotv\vec{g}\right)\vec{a}\\
 & =\vec{a}\dotv\text{\text{RHS}\ensuremath{\quad}}\forall\:\vec{a}\:.\:\Box
\end{align*}
\begin{align*}
\text{LHS}\dotv\text{\ensuremath{\vec{b}}} & =\vec{f}\cross\Ident\dotv\left(\vec{g}\cross\ensuremath{\vec{b}}\right)\\
 & =\vec{f}\cross\left(\vec{g}\cross\ensuremath{\vec{b}}\right)\\
 & =\left(\vec{f}\dotv\vec{b}\right)\vec{g}-\left(\vec{f}\dotv\vec{g}\right)\vec{b}\\
 & =\text{RHS}\dotv\text{\text{\ensuremath{\vec{b}}}\ensuremath{\quad}}\forall\:\vec{b}\:.\:\Box
\end{align*}

\section*{Acknowledgments}

We gratefully acknowledge useful discussions with Naoki Sato on constraint
options, Robert MacKay for suggesting the relevance of weak KAM theory
to 3-D MHD equilibrium theory, Joshua Burby for discussions of an
earlier version of this paper, Markus Hegland for a discussion of
regularization, Lindon Roberts for references on infinite-dimensional
augmented Lagrangian optimization methods and Matthew Hole for reading
the manuscript. We also thank Zoran Levnajić and Igor Mezić for consenting
to use of their visualizations of chaos in our Fig. \ref{fig:ErgodPartition}.
Finally we thank an anonymous referee for pointing out the interpretation
of the IOL Lagrange multiplier as an electrostatic polarization field.

The work of ZQ was supported by the Australian Research Council under
grant DP170102606 and the Simons Foundation/SFARI (560651, AB). RLD
and ZQ also acknowledge travel support from the Simons Foundation/SFARI
(560651, AB).

 \bibliographystyle{jpp}
\bibliography{RLDBibDeskPapers_MRXMHD}

\end{document}